\documentclass[aps,prd,10pt,a4paper,twocolumn,amsfonts,amsmath,amssymb,superscriptaddress,notitlepage,preprintnumbers]{revtex4-2}

\usepackage{hyperref}
\usepackage{bm}
\usepackage{graphicx}
\usepackage{comment}
\usepackage[dvipsnames]{xcolor}

\newcommand{\hMpcinv}{h\,\mathrm{Mpc}^{-1}}
\newcommand{\Pfid}{P^\mathrm{fid}}
\newcommand{\reg}{\mathrm{reg}}
\newcommand{\sigmad}{\sigma_\mathrm{d}}
\newcommand{\tar}{\text{tar}}
\newcommand{\fid}{\text{fid}}

\newcommand{\mnras}{Mon. Not. Roy. Astron. Soc.}

\newcommand{\jcap}{J. Cosmol. Astropart. Phys.}
\newcommand{\physrep}{Phys. Rept.}
\newcommand{\aap}{Astron. Astrophys.}

\newcommand{\pasj}{Publ. Astron. Soc. Jpn.}

\begin{document}

\preprint{YITP-21-71}

\title{Implementing spectra response function approaches for fast calculation of power spectra and bispectra}

\author{Ken Osato}
\email[]{ken.osato@yukawa.kyoto-u.ac.jp}
\affiliation{Center for Gravitational Physics, Yukawa Institute for Theoretical Physics,
Kyoto University, Kyoto 606-8502, Japan}
\affiliation{LPENS, D\'epartement de Physique, \'Ecole Normale Sup\'erieure,
Universit\'e PSL, CNRS, Sorbonne Universit\'e, Universit\'e de Paris, 75005 Paris, France}
\affiliation{Institut d'Astrophysique de Paris, Sorbonne Universit\'e, CNRS, UMR 7095, 75014 Paris, France}

\author{Takahiro Nishimichi}
\affiliation{Center for Gravitational Physics, Yukawa Institute for Theoretical Physics,
Kyoto University, Kyoto 606-8502, Japan}
\affiliation{Kavli Institute for the Physics and Mathematics of the Universe,
The University of Tokyo Institutes for Advanced Study (UTIAS), The University of Tokyo,
Chiba 277-8583, Japan}

\author{Atsushi Taruya}
\affiliation{Center for Gravitational Physics, Yukawa Institute for Theoretical Physics,
Kyoto University, Kyoto 606-8502, Japan}
\affiliation{Kavli Institute for the Physics and Mathematics of the Universe,
The University of Tokyo Institutes for Advanced Study (UTIAS), The University of Tokyo,
Chiba 277-8583, Japan}

\author{Francis Bernardeau}
\affiliation{Institut de physique th\'eorique, Universit\'e Paris Saclay CEA, CNRS, 91191 Gif-sur-Yvette, France}
\affiliation{Institut d'Astrophysique de Paris, Sorbonne Universit\'e, CNRS, UMR 7095, 75014 Paris, France}

\begin{abstract}
Perturbation theory of large-scale structures of the Universe
at next-to-leading order and next-to-next-to-leading order provides
us with predictions of cosmological statistics at sub-percent level in the mildly non-linear regime.
Its use to infer cosmological parameters from spectroscopic surveys, however, is hampered by the computational cost of making predictions
for a large number of parameters. In order to reduce the running time of the codes, we present a fast scheme in the context of
the regularized perturbation theory approach and applied it to power spectra at 2-loop level
and bispectra at 1-loop level, including the impact of binning.
This method utilizes a Taylor expansion of the power spectrum as a functional of
the linear power spectrum around fiducial points at which costly direct evaluation of perturbative diagrams
is performed and tabulated. The computation of the predicted spectra for arbitrary cosmological parameters
then requires only one-dimensional integrals that can be done within a few minutes. It makes
this method suitable for Markov chain Monte-Carlo analyses for cosmological parameter inference.
\end{abstract}

\date{\today}

%\maketitle must follow title, authors, abstract, \pacs, and \keywords
\maketitle

\section{Introduction}
The formation and evolution of large-scale structures (LSS) of the Universe
reflect primordial density perturbations generated from quantum fluctuations in an inflationary era,
gravitational evolution mainly driven by dark matter,
and accelerated expansion driven by dark energy.
Thus, the observations of LSS are the key to understand the physics in the early Universe
and the nature of the dark components (for a review, see Ref.~\onlinecite{Weinberg2013}).
The most fundamental statistics to characterize the density fluctuations is the
$n$-point correlation functions and their counterparts in Fourier space, i.e., poly-spectra.
In particular, since the density field is close to a Gaussian random field,
poly-spectra at the lowest orders contain a large fraction of information;
if the density field follows a Gaussian distribution, the
information content is completely described by its power spectrum.
A non-zero bispectrum, however, indicates a departure from Gaussian distribution,
sourced by primordial non-Gaussianity and/or non-linear gravitational evolution.
By analyzing these statistics, one can infer the energy budget of dark components and the geometry of the Universe
through the measurement of the baryon acoustic oscillation (BAO),
which provides a characteristic scale of
sound waves in primeval baryon and photon fluid \cite{Aubourg2015}
and is first detected through spectroscopic observations of galaxies by
2dF Galaxy Redshift Survey (2dFGRS; Ref.~\onlinecite{Cole2005})
and luminous red galaxies by Sloan Digital Sky Survey (SDSS; Ref.~\onlinecite{Eisenstein2005}).
The subsequent survey programs have measured the BAO scale more precisely with a wide range of redshifts:
6dF Galaxy Survey \cite{Beutrler2011},
WiggleZ \cite{Blake2011},
and Extended Baryon Oscillation Spectroscopic Survey (eBOSS) \cite{GilMarin2020,Bautista2021}.
Furthermore, upcoming survey programs aim to measure the statistics at an unprecedented precision
with wide and deep observations:
Dark Energy Spectroscopic Instrument (DESI) \cite{Martini2018},
Subaru Prime Focus Spectrograph (PFS) \cite{Takada2014},
\textit{Euclid} \cite{Amendola2018},
and Roman Space Telescope \footnote{\url{https://roman.gsfc.nasa.gov/}}.
In the analysis of the measurements, precise and accurate modeling
of cosmological statistics is essential to fully explore the potential of high-quality data provided
by the future surveys.

In the practical analysis,
perturbation theory (PT) of LSS is widely used as a standard approach
to predict cosmological statistics (for a review, see Ref.~\onlinecite{Bernardeau2002}).
By adopting the single stream approximation for cosmic fluid,
basic equations (continuity, Euler, and Poisson equations) can be solved in a perturbative manner,
which is referred to as standard perturbation theory (SPT).
The PT model is fully analytical and accurate at sub-percent level
in a fair range of scale for redshifts of interest
(roughly speaking, scales of $\lesssim 0.15, 0.20, 0.25 \, \hMpcinv$
at $z = 0.5, 1, 2$, respectively, for PT schemes at 2-loop level \cite{Taruya2012,Blas2014})
and the full shapes of the statistics, which are measured in wide redshift surveys, are predicted.
However, the capability of exploring the parameter space is
limited due to the computational cost of PT schemes.
This is the case for power spectra at next-to-next-to-leading order (NNLO)
and the situation is even more critical for bispectra even at next-to-leading order (NLO)
as a large number of possible triangle configurations of wave-vectors are required.

Different strategies have been advocated to circumvent these difficulties.
One line of investigation is represented by the effective field theory (EFT) approach \cite{Baumann2012,Carrasco2012,Carrasco2014},
where one hopes to capture the small scale physics with effective interaction terms in the fluid equations.
The EFT approach has been applied to the full shape analysis of galaxy power spectrum
measured by BOSS to infer cosmological parameters \cite{dAmico2020,Colas2020,Ivanov2020}.
A drawback of this method is that it makes use of a large number of
nuisance parameters
which can potentially degrade the constraining power of the method on cosmological parameters \cite{Osato2019}.

Means to reduce the computational costs have also been looked for with the use of specific computational schemes
\cite{Schmittfull2016a,Schmittfull2016b,McEwen2016,Fang2017,Simonovic2018}
for the computations of the loop-terms (based on methods such as FFTLog \cite{Hamilton2000}).
However, these fast schemes have been formulated and actually computed only for
the power spectrum at NLO so far and 2-loop level calculations have not yet been successfully implemented.
As more numerically oriented approach,
computational schemes calibrated with $N$-body simulations, which enable fast exploration,
have been proposed: fitting formulas based on halo model (\textit{Halofit}) \cite{Smith2003,Takahashi2012,Mead2015,Cataneo2019,Giblin2019},
response with respect to cosmological parameters \cite{Cataneo2017,Hannestad2020},
and emulator \cite{Heitmann2010,Heitmann2009,Lawrence2010,Nishimichi2019,EuclidCollaboration2019}.

In this work, we present an alternative scheme to compute power spectra at NNLO and
bispectra at NLO in a fast manner. The presentation of the method and its performance is made
in the context of regularized PT (RegPT) approach \cite{Taruya2012},
which reorganizes the SPT expansions with multi-point propagators \cite{Bernardeau2008}.
The method presented here is not however restricted to such a flavor of PT schemes.
In our scheme, the calculation of the relevant terms is accelerated
by introducing the spectra response function.
The response function is defined mathematically
through the functional derivative of the \textit{nonlinear} spectra
with respect to the linear power spectrum. In the case of power spectrum,
it quantifies the variation of the evolved power spectrum at a given mode
with respect to a small initial disturbance added in the linear power spectrum at other Fourier modes.
Making use of this function, the nonlinear power spectrum is Taylor-expanded
with respect to the linear power spectrum around a fiducial cosmological model.
Then, provided the response function data that is precomputed and is numerically tabulated,
one can compute the correction terms involving the response function
for a given linear power spectrum in a target cosmology.
Adding them to the fiducial (nonlinear) power spectrum,
the nonlinear power spectrum for a target cosmological model is reconstructed,
and the method gives an accurate prediction as long as the differences of the linear power spectrum
between the fiducial and target cosmological model is small.
In practice, preparing the precomputed data set in several fiducial models,
the reconstruction can be made more accurately,
making it possible to cover a wide range of cosmological parameter space.
The important point of this approach is that thanks to the statistical isotropy,
all the correction terms to be computed involve only the one-dimensional integrals.
This is true even at the higher-loop contributions.
Hence the prediction in this method can be made much faster than that in the direct method
involving higher-dimensional integrals.

The approach mentioned above was first proposed in Ref.~\cite{Taruya2012}
and was explicitly implemented in their PT (RegPT) code, named as RegPT-fast.
Later, the response function of the power spectrum was measured
in cosmological $N$-body simulation by Ref.~\cite{Nishimichi2016},
and the measured results has been compared with PT predictions.
Further, in Ref.~\cite{Nishimichi2017}, based on a high-precision measurement of the power spectrum
and response function, an accurate and fast template for nonlinear power spectrum has been developed.
The new template called \texttt{RESPRESSO} was shown to outperform existing PT templates,
and to be accurate even beyond the validity domain of PT calculations.
Also, physical interpretations of the measured response function,
that has been found to be suppressed for a coupling between large- and small-scale modes,
has been investigated in Ref.~\cite{Halle2020}.
Note here that these previous works have focused on the power spectrum.
However, the concept of response function is quite general,
and the method is applied to any statistics, including the bispectra.
As long as the statistical isotropy of initial condition holds,
the Taylor-expanded correction terms involves only the one-dimensional integrals,
and hence the approach can be used for accelerated statistical calculations even for higher-order statistics.
The aim of the present paper is to demonstrate this point
and to develop the fast PT templates for cross power spectra and bispectra of the density and velocity fields.

This paper is organized as follows.
In Section~\ref{sec:direct}, we overview PT approaches to compute power spectra and bispectra.
In Section~\ref{sec:fast}, we present the fast scheme
to compute the correction terms with functional derivatives.
In Section~\ref{sec:results}, we discuss the accuracy of the fast scheme
in comparison with the direct computation of the loop integrals
and then validate the calculations by comparing the results with $N$-body simulations.
We make concluding remarks in Section~\ref{sec:conclusions}.

Throughout this paper, we assume the flat $\Lambda$ cold dark matter (CDM) Universe.
Though our method is applicable for arbitrary cosmological parameters,
as a representative cosmology for validations of our method,
we adopt the best-fit cosmological parameters determined by the measurements of cosmic microwave background temperature and
polarization anisotropies of the \textit{Planck} mission \cite{Planck2015XIII}
(TT,TE,EE+lowP dataset). Hereafter, this cosmology is referred to as \textit{Planck} 2015.
$\omega_\mathrm{b} = \Omega_\mathrm{b} h^2 = 0.02225$,
$\omega_\mathrm{c} = \Omega_\mathrm{c} h^2 = 0.1198$, and $\omega_\nu = \Omega_\nu h^2 = 0.00064$ are
physical baryon, CDM, and massive neutrino densities, respectively,
where we assume that one of three generations of neutrinos has a finite mass of $0.06 \, \mathrm{eV}$
and the other generations are massless.
$\Omega_\Lambda = 0.6844$ is the density parameter of cosmological constant and
$h = H_0 / (100 \, \mathrm{km} \, \mathrm{s}^{-1} \, \mathrm{Mpc}^{-1})$
is the scaled Hubble parameter, which is determined through
$(\omega_\mathrm{b} + \omega_\mathrm{c} + \omega_\nu)/h^2 + \Omega_\Lambda = 1$.
$\log (10^{10} A_\mathrm{s}) = 3.094$ and $n_\mathrm{s} = 0.9645$ are
the amplitude and slope of primordial scalar perturbation, respectively,
at the pivot scale $k_\mathrm{piv} = 0.05 \, \mathrm{Mpc}^{-1}$.

\section{Direct Method}
\label{sec:direct}
In this Section, we review the SPT formalism and
multi-point propagator expansion based on RegPT approach \cite{Taruya2012}.
We refer to the naive numerical evaluation of these original expressions
as the ``direct'' method in contrast to the ``fast'' method
which will be presented in Section~\ref{sec:fast} because the direct method involves
direct integration up to five dimensions.

\subsection{Standard perturbation theory}
The framework of SPT relies on the single-stream treatment of the cosmological Vlasov--Poisson equation,
with which the governing equations are reduced to
those of the pressureless fluid systems (i.e., continuity, Euler, and Poisson equations).
Because our interest is in the late-time universe dominated by the growing-mode solutions,
the density and velocity fields are expanded in Fourier space as
\begin{equation}
  \Psi_a (\bm{k}, \eta) = \sum_{n=1}^\infty e^{n \eta} \Psi_a^{(n)} (\bm{k}) ,
  \label{eq:def_Psi_a}
\end{equation}
where $\Psi_a (\bm{k}, \eta) = (\delta (\bm{k}, \eta), - \theta (\bm{k}, \eta) / (a H f))$
is the density and velocity divergence doublet,
$\delta (\bm{k}, \eta)$ is the density contrast,
$\theta (\bm{k}, \eta)$ is the velocity divergence,
$a$ is the scale factor, $H (a)$ is the Hubble parameter,
$f \equiv \mathrm{d} \ln D_+ / \mathrm{d} \ln a$ is the growth rate,
and $D_+ (a)$ is the linear growth factor normalized as $D_+ (a = 1) = 1$.
Throughout the paper, we adopt the Einstein--de Sitter approximation, in which
the time dependence of higher-order perturbative corrections is
factorized in a similar manner to the Einstein--de Sitter case,
with the time dependence of the $n$-th order terms given by $e^{n \eta}$,
where $\eta$ id defined by $\eta\equiv \ln D_+ (a)$.
In Eq.~\eqref{eq:def_Psi_a}, the time-independent $n$-th order doublet,
$\Psi_a^{(n)}$, represents the mode coupling of the gravitational evolution
and is expressed in terms of the linear density field $\delta_0$:
\begin{align}
  \Psi_a^{(n)} (\bm{k}) =& \int \frac{\mathrm{d}^3 q_1 \cdots \mathrm{d}^3 q_n}{(2 \pi)^{3 (n-1)}}
  \delta_\mathrm{D} (\bm{k} - \bm{q}_1 - \cdots - \bm{q}_n)
  \nonumber \\
  & \times F^{(n)}_{a} (\bm{q}_1 , \ldots, \bm{q}_n) \delta_0 (\bm{q}_1) \cdots \delta_0 (\bm{q}_n) ,
  \label{eq:Psi_SPT}
\end{align}
where $F^{(n)}_{a}$ is the $n$-th order symmetrized kernel function,
$\delta_0 (\bm{k})$ is the linear density contrast in Fourier space,
and $\delta_\mathrm{D}$ is the Dirac delta function.
The explicit expressions for the perturbation theory kernels $F^{(n)}_a$ can be found
in the literature (e.g., Ref.~\onlinecite{Crocce2006}).

In this paper, we are interested in computing the auto- and cross-power spectra and bispectra
of the density and velocity-divergence fields,
which we respectively denote by  $P_{ab} (k;\eta)$ and bispectra $B_{abc} (k_1, k_2, k_3; \eta)$.
They are defined as
\begin{align}
  & \langle \Psi_a (\bm{k}; \eta) \Psi_b (\bm{k}'; \eta) \rangle
  \equiv (2\pi)^3 \delta_\mathrm{D} (\bm{k} + \bm{k}') P_{ab} (k; \eta), \\
  & \langle \Psi_a (\bm{k}_1; \eta) \Psi_b (\bm{k}_2; \eta) \Psi_c (\bm{k}_3; \eta) \rangle
  \nonumber \\
  & \equiv (2\pi)^3 \delta_\mathrm{D} (\bm{k}_1 + \bm{k}_2 + \bm{k}_3)
  B_{abc} (k_1, k_2, k_3; \eta) ,
\end{align}
where the subscripts, $a$, $b$ or $c$ represent
either density field $\delta$ or velocity-divergence field $\theta$.
Then, from the SPT expansion (Eqs.~\ref{eq:def_Psi_a} and \ref{eq:Psi_SPT}),
power spectrum and bispectra are expressed as loop expansion:
\begin{align}
  P_{ab}^\text{SPT} (k) =& P_{ab, \text{tree}} (k) + P_{ab, \text{1-loop}} (k) + \nonumber \\
  & P_{ab, \text{2-loop}} (k) + \cdots , \\
  B_{abc}^\text{SPT} (k_1, k_2, k_3) =& B_{abc,\text{tree}} (k_1, k_2, k_3) + \nonumber \\
  & B_{abc, \text{1-loop}} (k_1, k_2, k_3) + \cdots .
\end{align}
The tree level (leading-order) terms are given as
\begin{align}
  P_{ab, \text{tree}} (k) =& e^{2\eta} P_0 (k) , \\
  B_{abc,\text{tree}} (k_1, k_2, k_3) =& 2 e^{4 \eta} F^{(2)}_a (\bm{k}_2, \bm{k}_3) P_0 (k_2) P_0 (k_3)
  \nonumber \\
  & + \text{perms.} ,
\end{align}
where ``perms.'' is the sum of cyclic permutations of the first term
with respect to wave-vectors $(\bm{k}_1, \bm{k}_2, \bm{k}_3)$ and subscripts $(a, b, c)$.
The 1-loop level terms ($P_{ab, \text{1-loop}}$ and $B_{abc, \text{1-loop}}$) and
2-loop level term ($P_{ab, \text{2-loop}}$)
correspond to NLO and NNLO correction terms, respectively.
Note that the 1-loop corrections to the power spectrum are expressed
as the loop integrals involving square powers of the linear power spectrum and higher-order SPT kernels.
For the 2-loop corrections to the power spectrum and 1-loop corrections to the bispectrum,
the loop integrals involve cubic powers of the linear power spectrum.
It is to be noted that all these corrections are re-expressed in terms of the multi-point propagators
that are perturbatively evaluated with the SPT kernels.
Their explicit expressions for power spectra at NNLO and bispectra at NLO based on SPT
are found in Appendix~\ref{sec:expressions_SPT}.

\subsection{Multi-point propagator}
Next, let us consider the RegPT treatment as one of the resummed PT frameworks.
This treatment relies on the expansion
based on multi-point propagators, which are
the response of cosmic fields with respect to the initial fields
and originally proposed by Ref.~\cite{Bernardeau2008}.
The SPT expansion can be reorganized with multi-point propagators
and since nonperturbative properties are incorporated in
multi-point propagtors through taking infinite sum of the SPT expansion,
the RegPT treatment achieves better convergence \footnote{It is known that
resummed PT such as RegPT breaks the generalized Galilean invariance
(as pointed in \cite{2013arXiv1311.2724B} and more recently in \cite{Peloso2017}).
In case of RegPT, the breaking of this invariance happens at order $k^6 \sigmad^6$,
where $\sigmad$ is ultimately calibrated against $N$-body simulations.
It happens therefore at scales where results of PT schemes are not reliable anymore.
The damping factor introduced in RegPT,
although it breaks the Galilean invariance,
eventually ensures that such unreliable scales are damped away in particular
in the computation of correlation functions in configuration space (as shown in \cite{Taruya2013}).
}.

First, we introduce the $(n+1)$-point propagator of the density and velocity divergence fields
through their functional derivative:
\begin{align}
  & \frac{1}{n!} \left\langle
  \frac{\delta^n \Psi_a (\bm{k} ; \eta)}{\delta \delta_0 (\bm{k}_1) \cdots \delta \delta_0 (\bm{k}_n)}
  \right\rangle \nonumber \\
  & \equiv \delta_\mathrm{D} (\bm{k} - \bm{k}_{1 \ldots n}) \frac{1}{(2 \pi)^{3(n-1)}}
  \Gamma_a^{(n)} (\bm{k}_1, \ldots, \bm{k}_n ; \eta) ,
\end{align}
where $\bm{k}_{1 \ldots n} \equiv \bm{k}_1 + \cdots + \bm{k}_n$.
Applying the SPT expansion, the $\Gamma$ function, which we call the propagator,
is perturbatively evaluated, with each building block involving the linear power spectrum and the SPT kernels:
\begin{align}
  & \Gamma_a^{(n)} (\bm{k}_1, \ldots, \bm{k}_n ; \eta)
  \nonumber \\
  & = \Gamma^{(n)}_{a,\text{tree}} (\bm{k}_1, \ldots, \bm{k}_n ; \eta ) +
  \sum_{p = 1}^{\infty} \Gamma_{a, p\text{-loop}}^{(n)} (\bm{k}_1, \ldots, \bm{k}_n  ; \eta ) , \\
  & \Gamma^{(n)}_{a,\text{tree}} (\bm{k}_1, \ldots, \bm{k}_n ; \eta ) \equiv
  e^{n \eta} F^{(n)}_a (\bm{k}_1, \ldots, \bm{k}_n) , \\
  & \Gamma_{a, p\text{-loop}}^{(n)} (\bm{k}_1, \ldots, \bm{k}_n ; \eta )
  \nonumber \\
  & \equiv e^{(n+2p) \eta} c_p^{(n)} \int \frac{\mathrm{d}^3 q_1 \ldots \mathrm{d}^3 q_p}{(2 \pi)^{3p}}
  \nonumber \\
  & \times F^{(n+2p)}_a (\bm{q}_1, -\bm{q}_1, \ldots, \bm{q}_p, -\bm{q}_p,
  \bm{k}_1, \ldots, \bm{k}_n)
  \nonumber \\
  & \times P_0 (q_1) \cdots P_0 (q_p) , \\
  & \equiv e^{(n+2p) \eta} \bar{\Gamma}_{a, p\text{-loop}}^{(n)} (\bm{k}_1, \ldots, \bm{k}_n),
\end{align}
where $c_p^{(n)} \equiv {}_{(n+2p)} C_n (2p-1)!!$
and $P_0 (k)$ is the linear power spectrum normalized at the present time.
It is to be noted that taking the high-$k$ limit, higher-order contributions can be systematically computed
at all orders, and as a result of summing up all the contributions, the multi-point propagators are shown to be exponentially suppressed as follows:
\begin{align}
\label{eq:Gamma_high-k}
  & \lim_{k \to \infty} \Gamma^{(n)}_a (\bm{k}_1, \ldots, \bm{k}_n ; \eta )
  \nonumber \\
  & = \exp \left[ -\frac{k^2 e^{2\eta} \sigmad^2}{2} \right]
  \Gamma^{(n)}_{a,\text{tree}} (\bm{k}_1, \ldots, \bm{k}_n ; \eta ) ,
\end{align}
where $k \equiv |\bm{k}_{1 \ldots n}|$.
The characteristic scale of the damping is determined by
the root-mean-square of the displacement field $\sigmad$:
\begin{equation}
  \label{eq:sigma_d}
  \sigmad^2 (k) = \int_0^{k_\Lambda (k)} \frac{\mathrm{d} q}{6 \pi^2} P_0 (q) ,
\end{equation}
where $k_\Lambda (k)$ is the UV cutoff scale.
In Ref.~\cite{Taruya2012}, comparing the predictions of power spectra with $N$-body simulations,
a running UV cutoff of $k_\Lambda (k) = k/2$ was adopted.
We adopt the same UV cutoff for power spectra calculations
but in order to achieve better match with simulations for bispectra,
a different cutoff scale $k_\Lambda (k) = k/6$ is employed for bispectra calculations.
In Appendix~\ref{sec:UV_cutoff}, we address how the UV cutoff scale affects
power spectra and bispectra.
As it has been advocated by Ref.~\cite{Bernardeau2012}
the exponential damping can be interpreted as the result of resummation of diagrams in a
hard part (high-$k$), and the
displacement dispersion $\sigmad$ in the exponent must be evaluated
in a consistent way that the domain of the integral is restricted to a soft part (low-$k$).
This implies that the boundary of the UV and
IR domains can be changed, depending
on not only the scale but also the quantity of our interest.

Then, the power spectra and bispectra can be expressed as infinite series
of contributions from propagators:
\begin{align}
  P_{ab} (k; \eta) =& \sum_{t=1}^{\infty} t! \int \frac{\mathrm{d}^3 q_1 \cdots \mathrm{d}^3 q_t}{(2 \pi)^{3(t-1)}}
  \delta_\mathrm{D} (\bm{k} - \bm{q}_{1 \ldots t}) \nonumber \\
  & \times \Gamma_a^{(t)} (\bm{q}_1, \ldots , \bm{q}_t ;\eta)
  \Gamma_b^{(t)} (\bm{q}_1, \ldots , \bm{q}_t ;\eta) \nonumber \\
  & \times P_0 (q_1) \cdots P_0 (q_t) ,
\end{align}
\begin{widetext}
\begin{align}
  \langle \Psi_a (\bm{k}_1; \eta) \Psi_b (\bm{k}_2; \eta) \Psi_c (\bm{k}_3; \eta) \rangle =&
  \sum_{r,s,t}^{\infty} \binom{r+s}{r} \binom{s+t}{t} \binom{t+r}{t} r! s! t!
  \int \frac{\mathrm{d}^3 q_1 \cdots \mathrm{d}^3 q_r}{(2 \pi)^{3(r-1)}}
  \frac{\mathrm{d}^3 q'_1 \cdots \mathrm{d}^3 q'_s}{(2 \pi)^{3(s-1)}}
  \frac{\mathrm{d}^3 q''_1 \cdots \mathrm{d}^3 q''_t}{(2 \pi)^{3(t-1)}}
  \nonumber \\
  & \times \delta_\mathrm{D} (\bm{k}_1 - \bm{q}_{1 \ldots r} - \bm{q}'_{1 \ldots s})
  \delta_\mathrm{D} (\bm{k}_2 + \bm{q}'_{1 \ldots s} - \bm{q}''_{1 \ldots t})
  \delta_\mathrm{D} (\bm{k}_3 + \bm{q}''_{1 \ldots t} + \bm{q}_{1 \ldots r})
  \nonumber \\
  & \times \Gamma_a^{(r+s)} (\bm{q}_1, \ldots , \bm{q}_r, \bm{q}'_1, \ldots , \bm{q}'_s ;\eta)
  \Gamma_b^{(s+t)} (-\bm{q}'_1, \ldots , -\bm{q}'_s, \bm{q}''_1, \ldots , \bm{q}''_t ;\eta)
  \nonumber \\
  & \times \Gamma_c^{(t+r)} (-\bm{q}''_1, \ldots , -\bm{q}''_t, -\bm{q}_1, \ldots , \bm{q}_r ;\eta)
  \nonumber \\
  & \times P_0 (q_1) \cdots P_0 (q_r) P_0 (q'_1) \cdots P_0 (q'_s) P_0 (q''_1) \cdots P_0 (q''_t) ,
\end{align}
\end{widetext}
where the summation runs for non-negative integers except $r = s = t = 0$ in the latter equation.
In practice, we have to truncate the series at an appropriate order.
Nevertheless, the multi-point propagators as building blocks of the PT expansion are
fully non-perturbative objects, and include infinite series of higher-order corrections in the SPT expansion.
Provided an accurate construction of propagators that consistently interpolate between SPT results at low-$k$
and the resummed behavior at high-$k$ (Eq.~\ref{eq:Gamma_high-k}),
the above expansions are expected to deliver an improved prediction
for the power spectrum and bispectrum even truncating the series at finite order.
In this paper, similarly to the SPT case,
we consider the multi-point propagator expansion of the power spectra at 2-loop order and
of the bispectra at 1-loop order, whose expressions,
together with the construction of \textit{regularized} propagators,
are presented in Appendix~\ref{sec:expressions}.

\section{Fast Method}
\label{sec:fast}

\subsection{Taylor Expansion with Functional Derivatives}
As we saw in the previous section, the expressions of the power spectra
at 2-loop order involve six-dimensional integrals in both SPT and RegPT.
Although the statistical isotropy ensures that these are reduced to the five-dimensional integrals,
the remaining integrals are not easy to analytically perform.
Thus, we need a numerical technique to compute multi-dimensional integrals,
which is generally time-consuming.
This could be a critical bottleneck when using the SPT or RegPT power spectra as a theoretical template
in the Bayesian inference of cosmological parameters with the Markov chain Monte-Carlo (MCMC) technique.
For bispectra, the expressions involve three-dimensional integrals at the highest,
but one has to compute them at many possible configurations of wave-vector triangles.
Therefore, reducing the computational cost is an important task both for the power spectrum and bispectrum
to fully explore the parameter space.
In order to accelerate the PT calculations, in this paper,
we develop the response function approach,
which is similar to fast-RegPT introduced by Refs.~\cite{Taruya2012,Nishimichi2017}.
We first precompute the power spectra and bispectra for given sets of cosmological parameters,
and Taylor-expand the spectra taking the functional derivative with respect to the linear power spectrum.
Then, we add the leading-order correction term
to the precomputed spectra to account for the cosmology dependence.
The Taylor expansions of power spectra and bispectra are given as
\begin{widetext}
\begin{align}
  P_{ab}^\tar (k; \bm{\theta}^\tar, \sigmad^\tar) =&
  P_{ab}^\fid (k; \bm{\theta}^\fid, \sigmad^\tar)
  + \sum_{n=1}^\infty \frac{1}{n!} \int \mathrm{d} q_1 \cdots \mathrm{d} q_n
  \frac{\delta^n P_{ab}[P_0; \bm{\theta}^\fid, \sigmad^\tar]}{\delta P_0 (q_1) \cdots \delta P_0 (q_n)}
  \delta P_0 (q_1) \cdots \delta P_0 (q_n),
  \label{eq:rec_pk}
  \\
  B_{abc}^\tar (\bm{k}_1, \bm{k}_2, \bm{k}_3; \bm{\theta}^\tar, \sigmad^\tar) =&
  B_{abc}^\fid (\bm{k}_1, \bm{k}_2, \bm{k}_3; \bm{\theta}^\fid, \sigmad^\tar)
  + \sum_{n=1}^\infty \frac{1}{n!} \int \mathrm{d} q_1 \cdots \mathrm{d} q_n
  \frac{\delta^n B_{abc}[P_0; \bm{\theta}^\fid, \sigmad^\tar]}{\delta P_0 (q_1) \cdots \delta P_0 (q_n)}
  \delta P_0 (q_1) \cdots \delta P_0 (q_n),
  \label{eq:rec_bk}
\end{align}
\end{widetext}
where $\bm{\theta}$ is a set of cosmological parameters
and $\delta P_0 (k) \equiv P_0^\tar (k) - P_0^\fid (k)$.
The superscripts ``tar'' and ``fid'' denote a target cosmological model at which we wish to make a prediction,
and a fiducial cosmological model chosen from the precomputed models, respectively.
Note that the rms of displacement field $\sigma_\mathrm{d}$
in the fiducial spectra and functional derivatives,
are evaluated for the target model instead of the fiducial model
as in the standard Taylor-series expansion,
because the dependence of this parameter can be incorporated
fully analytically without performing a new direct evaluation of the loop integrals at the target model.
If the difference of the linear spectra $\delta P_0 (k)$ between the target and the fiducial model is small enough,
the expansion truncated up to the leading order correction (first terms in the infinite series in Eqs.~\ref{eq:rec_pk} and \ref{eq:rec_bk})
gives reasonable estimates of the target spectra (see Section~\ref{sec:validation}).
In contrast, if the difference is large, higher order contributions, e.g.,
terms beyond the first one in the infinite series in Eqs.~\eqref{eq:rec_pk} and \eqref{eq:rec_bk},
are not negligible.
The explicit expressions and implementation of the correction terms,
which involve up to one-dimensional integrals, are presented in Appendix~\ref{sec:kernels}.
The functional derivatives have to be evaluated only once at every fiducial model,
and for arbitrary target cosmology, we compute only one-dimensional
integrals to obtain the full spectra.
The choice of the fiducial models is however not fully straightforward
and has to be made in an efficient manner as explained in what follows.

\subsection{Fiducial models and the distance metric}
\label{sec:distance}
We have discussed that a fast evaluation of the spectra can be performed by starting from a fiducial model
for which all the necessary terms are already precomputed.
Ideally the technique we advocate will be all the more accurate that one can choose fiducial models that are as close to the target models as possible.
We then need to prepare multiple fiducial models to cover the cosmological parameter space,
and an efficient arrangement of these models are crucial for efficiency,
both in terms of the computational cost of the kernel diagrams at the fiducial model locations
and in terms of the size of the numerical tables to store the precomputed data.

Before explaining our procedure in details, note that
a simple trick can be used to improve the efficiency dramatically. This relies on the fact that
all the terms in Eqs.~\eqref{eq:rec_pk} and \eqref{eq:rec_bk}
are factorizable with respect to the overall amplitude of the linear power spectrum
and can therefore freely be chosen given the target model.
Namely, in order to reduce the difference between the fiducial and target power spectra,
a scaling factor $c$ is introduced:
\begin{equation}
  P^\mathrm{fid}_0 (k) \to c P^\mathrm{fid}_0 (k).
\end{equation}
The factor $c$ can be chosen to minimize a distance metric as discussed below.

Though the choice of fiducial cosmologies is arbitrary,
it is important to choose a set of fiducial cosmologies
which are homogeneously sampled and efficiently cover the large volume of parameter space.
For this purpose, we employ the latin hypercube design
(for a review, see Ref.~\onlinecite{Garud2017})
to define fiducial cosmological parameters.
As a working example, we select 10 fiducial cosmologies
and 10 target cosmologies for validation test.
We use a public R package SLHD~\cite{SLHD},
which can generate ``slices'' of samples optimized both within and across the slices.
In our case, the 10 fiducial and the 10 validation samples are taken from 2 independent slices,
and the full 20 model design is also optimized to evenly sample the parameter space.
The range of parameters are determined following the setup in Ref.~\cite{Nishimichi2019}:
\begin{align}
  0.211375 < \omega_\mathrm{b} < 0.0233625, \nonumber \\
  0.10782 < \omega_\mathrm{c} < 0.13178, \nonumber \\
  0.54752 < \Omega_\mathrm{de} < 0.82128, \nonumber \\
  0.916275 < n_\mathrm{s} < 1.012725 .
  \label{eq:range}
\end{align}
Note that the expressions are always factorizable for the amplitude of linear power spectra
so the dependence on the primordial scalar amplitude parameter $A_\mathrm{s}$ can be incorporated
in an exact manner. In practice we fix $\log (10^{10} A_\mathrm{s}) = 3.094$ for all fiducial models.

\begin{figure*}
  \includegraphics[width=0.9\textwidth]{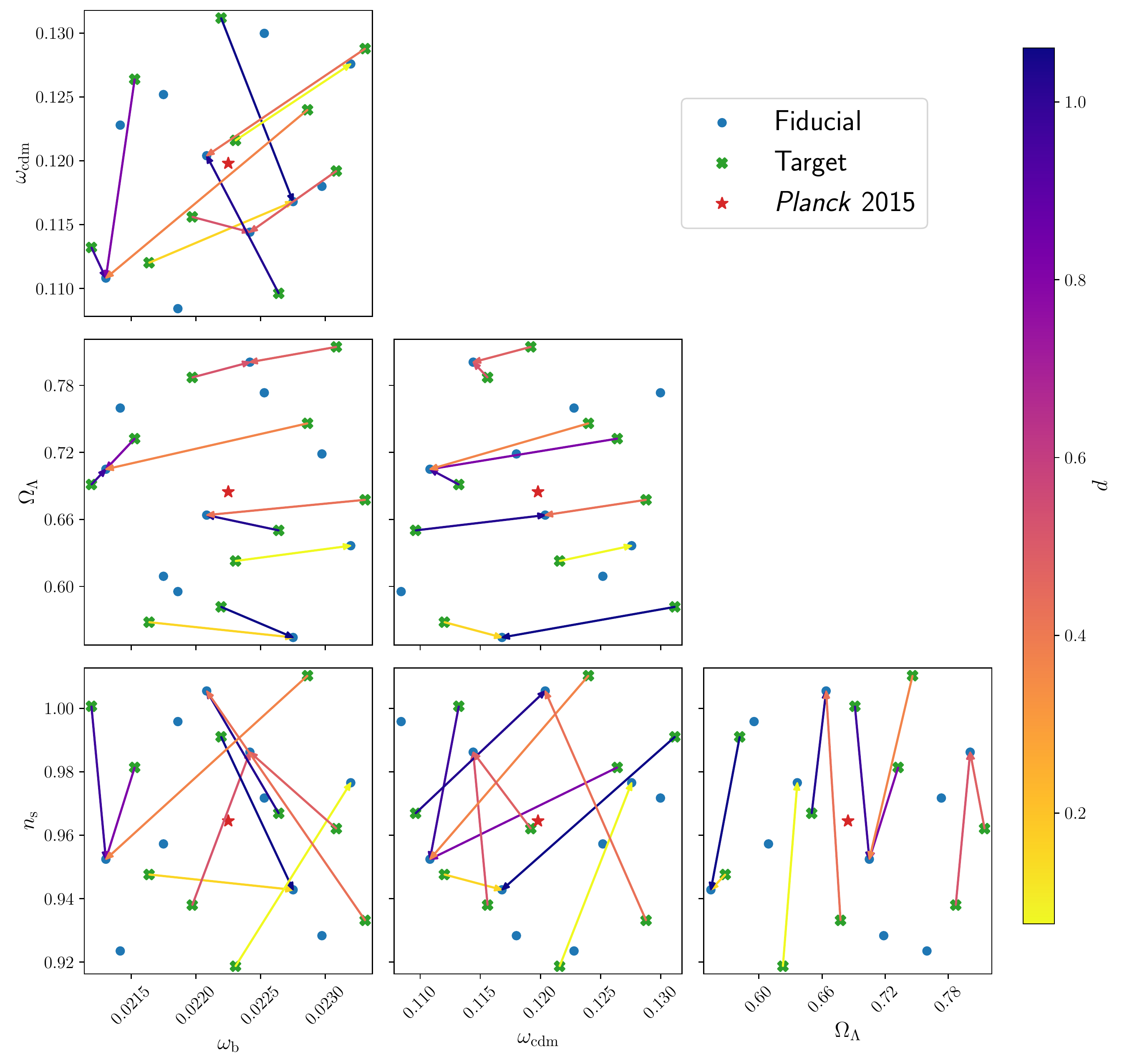}
  \caption{The fiducial and target cosmological parameters.
  The arrow from each target cosmology
  indicates the nearest fiducial cosmology
  and its color corresponds to the distance (Eq.~\ref{eq:distance}).}
  \label{fig:design}
\end{figure*}

The scaling factor $c$ is then determined in the following way.
First, we define the distance between cosmologies:
\begin{equation}
  \label{eq:distance}
  d^2 = \frac{1}{n_k} \sum_{i = 1}^{n_k}
  \frac{\left[ \log P^\mathrm{tar}_0 (k_i) - \log \left( c P^\mathrm{fid}_0 (k_i) \right) \right]^2}
  {\sigma_{k_i}^2} ,
\end{equation}
where $\sigma_{k_i} = k_i/k_0$, $k_0 = 1 \, h \, \mathrm{Mpc}^{-1}$, and $n_k = 20$.
For each fiducial cosmology, we determine the scaling factor which minimizes the square sum of the distance
at logrithmically equal-spaced $k$-bins in the range $[0.15, 1.0]\,h\,\mathrm{Mpc}^{-1}$.
The choice of the wave-numbers is made to absorb the difference of the power spectra mainly on small scales
with the scaling factor.
Then, we recompute the distance with the scaling factor optimized in the previous step, but now at other
logrithmically equal-spaced $k$-bins between $[0.01, 1.0]\,h\,\mathrm{Mpc}^{-1}$
to cover a wider wavenumber range.
These bins are used to find the fiducial model which minimizes the distance.
With these procedures, we can determine the nearest fiducial model
and the corresponding scaling factor for each target model.
This is illustrated in Figure~\ref{fig:design}, where the fiducial and validation models are
depicted by circles and crosses, respectively.
The fiducial-validation model pairs are indicated by the straight lines connecting them,
with the color showing the distance metric.

\subsection{Sampling and interpolation}
As one of the technical remarks,
we here discuss the sampling and interpolation of the results obtained from the fast method.
First note that the power spectra, bisectra and precomputed kernels
in fiducial cosmological models are stored as numerical table for fixed wave-vectors.
In particular, the sampling points of kernels are determined
in an optimal way to perform numerical integration
with the Gaussian quadrature.
Thus, in order to obtain the target-model predictions
at arbitrary wave-vectors, we need an interpolation.
In doing so, one strategy is to first build the full data set for the target model sampled
at the same wave-vectors as those in fiducial models.
We then use this as precomputed results
to get the predictions of power spectra and bispectra
at arbitrary Fourier modes based on the interpolation.
With this approach, the reconstruction of the spectra which requires the numerical integration needs only once,
and one can quickly evaluate power spectra
and bispectra for large number of wave-vectors and triangles,
thus speeding up the calculations.
This approach is particularly suited for the cosmological parameter estimation with MCMC technique and
is critical to compute the binned bispectrum (see Section~\ref{sec:comp_sims}).

Here we make comments on the sampling scheme of wave-vectors adopted in both fiducial and target models.
For power spectrum, the wave-vector is log-equally sampled in the range $[10^{-3}, 1] \, \hMpcinv$ with the number of sampling $n_k = 120$.
Then, for arbitrary wave-vector,
we employ cubic spline with the precomputed results.
For bispectrum, in order to reduce the dynamic range, we apply interpolation for reduced bispectra $Q(k_1, k_2, k_3)$:
\begin{align}
  & Q(k_1, k_2, k_3) \equiv
  \nonumber \\
  & \frac{B(k_1, k_2, k_3)}
  {P_\mathrm{L}(k_1) P_\mathrm{L}(k_2) + P_\mathrm{L}(k_2) P_\mathrm{L}(k_3)
  + P_\mathrm{L}(k_3) P_\mathrm{L}(k_1)},
\end{align}
where $P_\mathrm{L} (k)$ is the linear power spectrum at the same redshift as the bispectrum.
Since the sampling in $(k_1, k_2, k_3)$ space is subject to the triangle condition,
we introduce a different coordinate system $(K_1, K_2, K_3)$ for uniform sampling,
which is defined as
\begin{equation}
\label{eq:def_K}
  \begin{pmatrix}
    K_1 \\
    K_2 \\
    K_3
  \end{pmatrix}
  =
  \begin{pmatrix}
    k_2 + k_3 - k_1 \\
    k_3 + k_1 - k_2 \\
    k_1 + k_2 - k_3
  \end{pmatrix}
  .
\end{equation}
A positive $K_i \, (i = 1, 2, 3)$ ensures the triangle condition.
As our default setting,
we apply log-equally sampling with respect to $(K_1, K_2, K_3)$
in the range $[10^{-3}, 0.6] \, h \, \mathrm{Mpc}^{-1}$
with the number of sampling $n_K = 100$ in each dimension,
and $n_K^3 = 10^6$ evaluations of bispectra with the fast method are performed as precomputation.
Once the precomputation is done,
bispectrum for an arbitrary triangle is calculated with linear interpolation of the precomputed bispectra.
We have confirmed that the interpolation is generally accurate within $0.1 \%$ but can become worse as $1\%$
if one of $K_i$ is large (i.e., $K_i \simeq 0.5 \text{--} 0.6 \, h \, \mathrm{Mpc}^{-1}$).
If the bispectrum outside this range is requested,
we switch to the tree-level bispectrum.

\section{Results}
\label{sec:results}

\subsection{Validations of the fast scheme}
\label{sec:validation}
In this section, we present the performance of our fast scheme.
We set the 10 models in the validation set as the target models,
for which we compute the power spectra and bispectra
both with the direct method and with the fast scheme.
Since the target cosmological models are sampled from an optimized latin hypercube design
with the distance to the nearest fiducial model available for the reconstruction maximized,
we expect that the deviation between the direct and fast methods roughly at these samples
corresponds to the inaccuracy at the worst cases and, in practice,
the accuracy is expected to be better at parameter points randomly chosen from the range in Eq.~\eqref{eq:range}.
All the results discussed in the main text are evaluated at the redshift $z = 1$
and results for other redshifts are presented in Appendix~\ref{sec:other_z}.

In Figure~\ref{fig:Pk_validation}, we show density auto-, density-velocity cross-, and
velocity-auto power spectra for the 10 target models.
The difference between the direct and fast results mildly depends on the wave-vector;
the accuracy degrades towards small scales for all the three spectra.
Furthermore, the accuracy is not severely affected by the distance between the target
and the nearest fiducial model, which is indicated by the darkness of the color of the lines.
In general, the accuracy is better than $0.5\%$ up to $k = 0.3 \, h \,\mathrm{Mpc}^{-1}$,
which assures reasonable performance for realistic measurements.

\begin{figure*}
  \includegraphics[width=\textwidth]{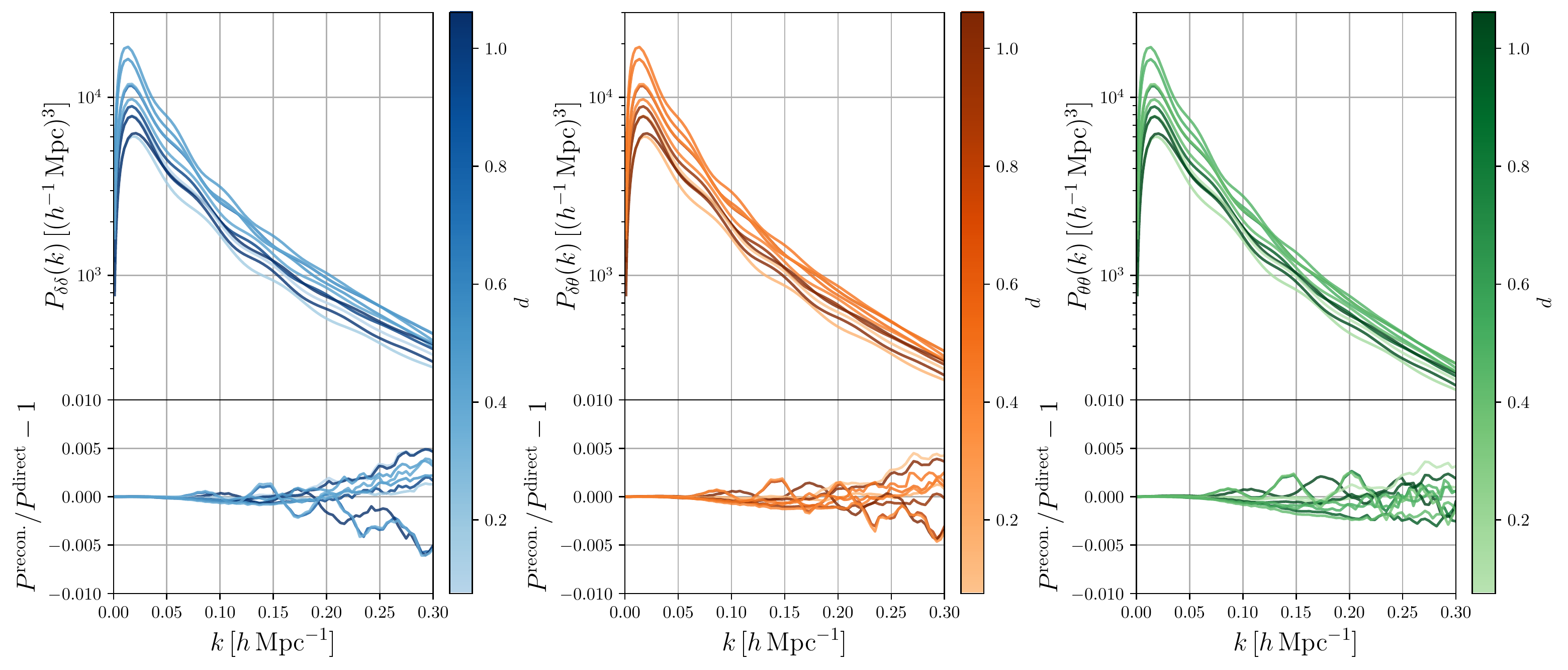}
  \caption{The reconstructed density auto-, density-velocity cross-,
  velocity auto-power spectra at the redshift $z = 1$.
  In the upper panels, we show power spectra of 10 target models computed with the fast scheme and
  the color corresponds to the distance between the target and the nearest fiducial models.}
  \label{fig:Pk_validation}
\end{figure*}

\begin{figure*}
  \includegraphics[width=\textwidth]{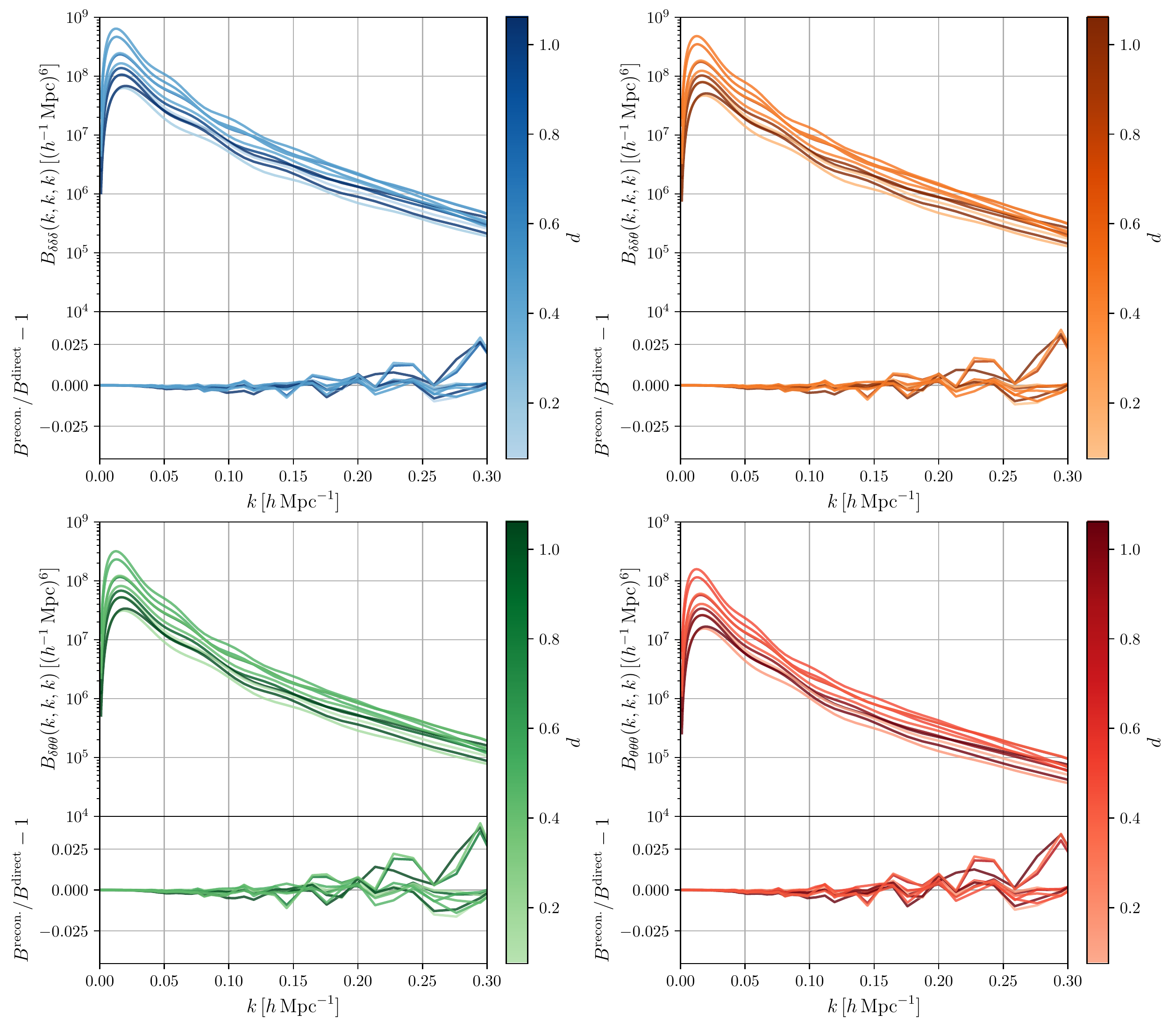}
  \caption{The reconstructed and direct bispectra for equilateral configuration
  with all combinations of density and velocity at the redshift $z = 1$.
  In the upper panels, we show power spectra of 10 target models computed with the fast scheme and
  the color corresponds to the distance between the target and the nearest fiducial models.}
  \label{fig:Bk_equil_validation}
\end{figure*}

\begin{figure*}
  \includegraphics[width=\textwidth]{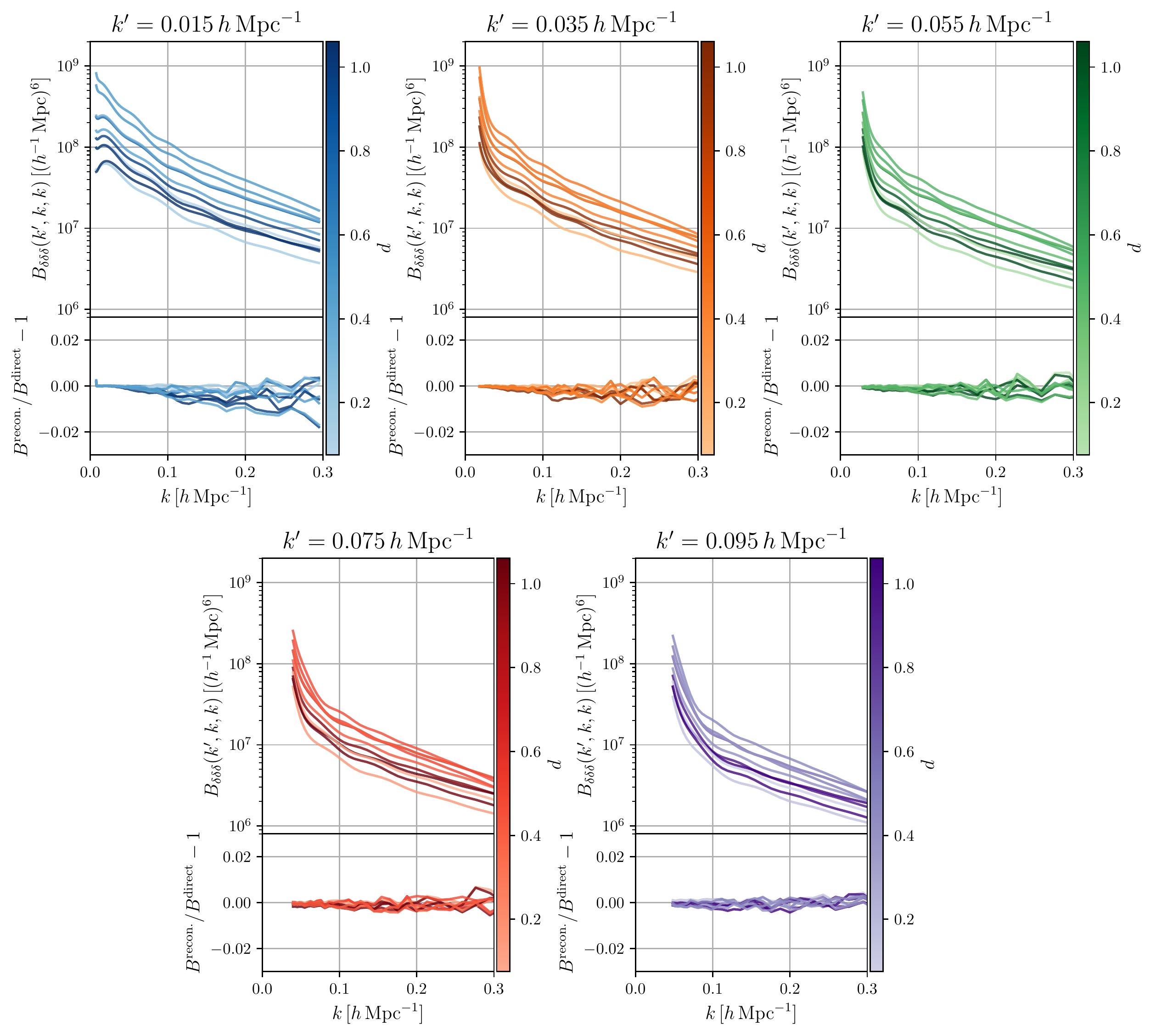}
  \caption{The reconstructed and direct density bispectra for isosceles configurations $(k', k, k)$
  with $k'/(h \, \mathrm{Mpc}^{-1}) = 0.015, 0.035, 0.055, 0.075, 0.095$ at the redshift $z = 1$.
  In the upper panels, we show power spectra of 10 target models computed with the fast scheme and
  the color corresponds to the distance between the target and the nearest fiducial models.}
  \label{fig:Bk_iso1_validation}
\end{figure*}

\begin{figure*}
  \includegraphics[width=\textwidth]{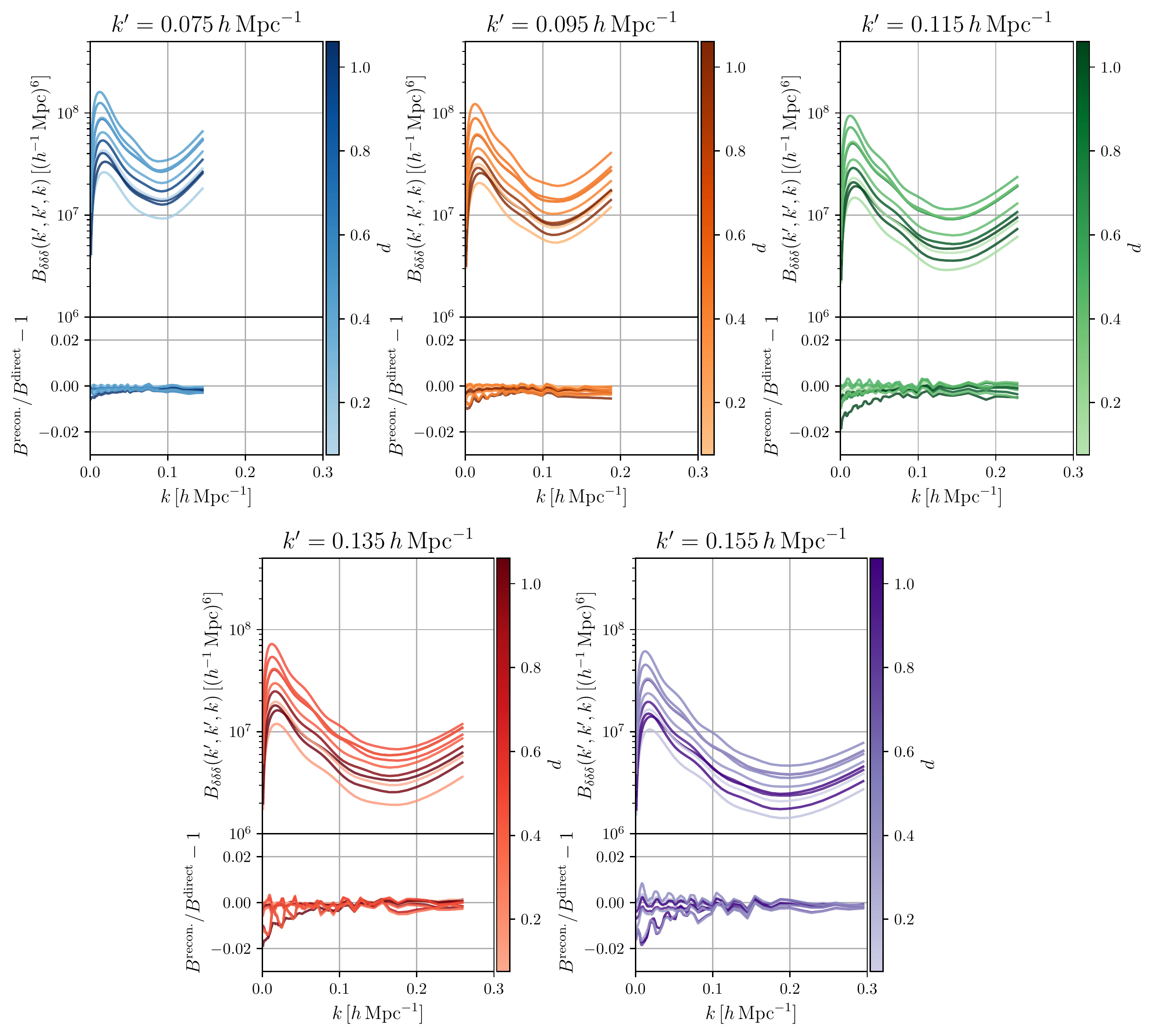}
  \caption{The reconstructed and direct density bispectra for isosceles configurations $(k', k', k)$
  with $k'/(h \, \mathrm{Mpc}^{-1}) = 0.075, 0.095, 0.115, 0.135, 0.155$ at the redshift $z = 1$.
  In the upper panels, we show power spectra of 10 target models computed with the fast scheme and
  the color corresponds to the distance between the target and the nearest fiducial models.}
  \label{fig:Bk_iso2_validation}
\end{figure*}

Next, we show the results of bispectra for equilateral configurations
with all possible combinations of density and velocity
in Figure~\ref{fig:Bk_equil_validation} and for two sequences of isosceles configurations:
$(k', k, k)$ in Figure~\ref{fig:Bk_iso1_validation} and
$(k', k', k)$ in Figure~\ref{fig:Bk_iso2_validation} for five fixed values of $k'$.
For equilateral configurations, the accuracy is within $2.5\%$
at $k = 0.3 \, h \, \mathrm{Mpc}^{-1}$
and almost similar for all combinations of density and velocity.
For isosceles configurations, the accuracy becomes slightly better and
is within $2\%$ at $k = 0.3 \, h \, \mathrm{Mpc}^{-1}$.
We only show the density bispectra for isosceles configurations but
the performance for any other combinations of density and velocity
is similar as shown in the case of equilateral configurations.

\subsection{Comparison with $N$-body simulations and binning effect}
\label{sec:comp_sims}
Here, we compare our PT results with power spectra and bispectra
measured from $N$-body simulations.
The details of $N$-body simulations used here are described in Ref.~\cite{Nishimichi2017}.
In this section, we show the results at the redshift $z = 0.901$,
which is the closest to the redshift $z = 1$ among the snapshots of the $N$-body simulations
presented in the reference and the results for other redshifts are presented in Appendix~\ref{sec:other_z}.
Since the measurement of the velocity divergence requires an interpolation scheme
as velocities are defined rigorously only at the exact locations of the simulation particles,
the results for the spectra involving only the density field are presented.

\begin{figure}
  \includegraphics[width=\columnwidth]{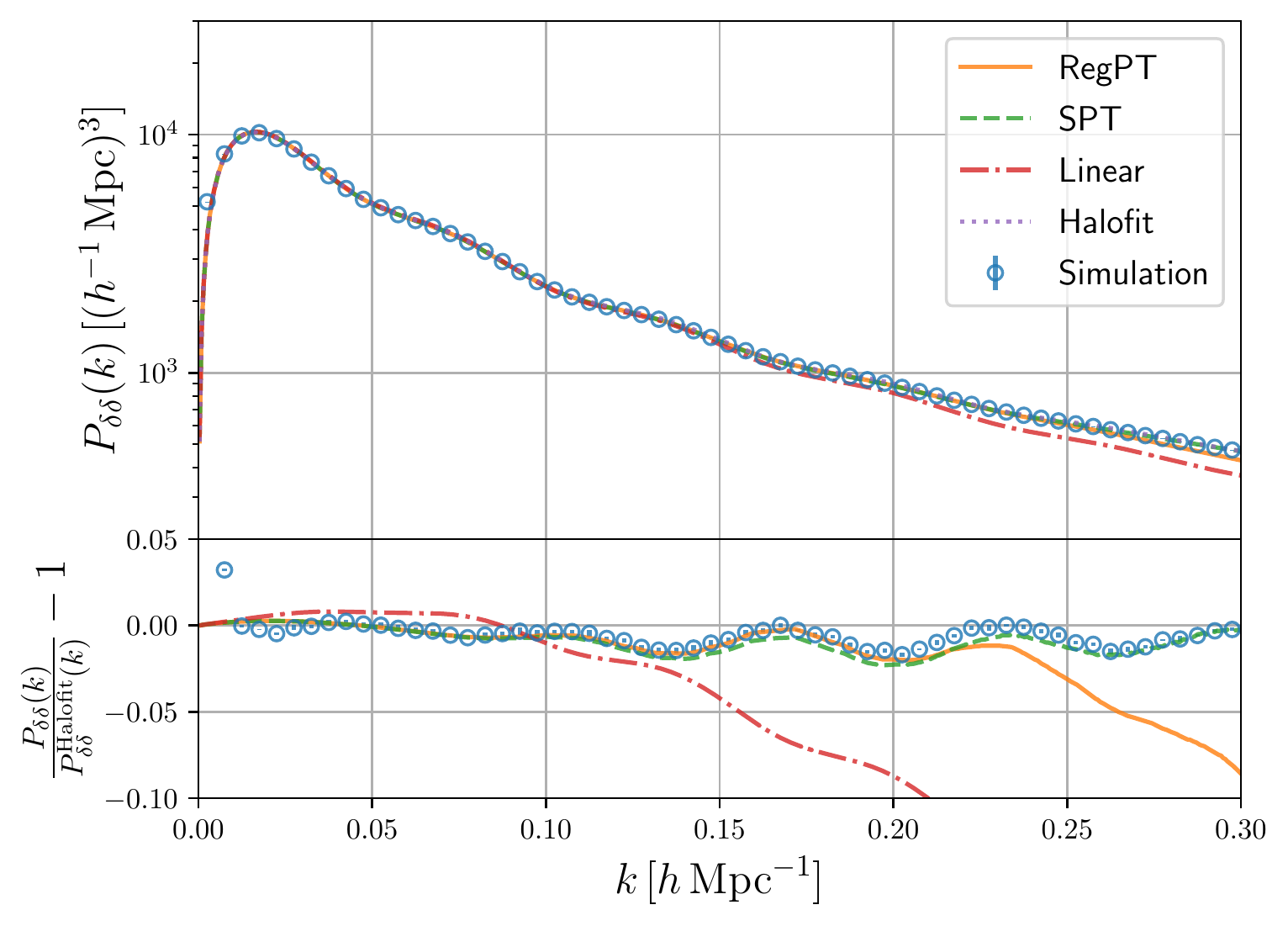}
  \caption{Comparison with power spectra measured from $N$-body simulation,
  linear power spectrum, predictions based on SPT and RegPT,
  and the \textit{Halofit} fitting formula at the redshift $z = 0.901$.
  In the lower panel, the ratios in terms of \textit{Halofit} are shown.}
  \label{fig:Pk}
\end{figure}

In Figure~\ref{fig:Pk}, we show the comparison between the power spectrum measured from $N$-body simulations,
the linear power spectrum, PT predictions based on SPT and RegPT,
and the fitting formula \textit{Halofit} \cite{Takahashi2012}.
At mildly non-linear regime (up to $k \simeq 0.2 \, \hMpcinv$),
SPT and RegPT give accurate predictions but RegPT slightly outperforms SPT.
At small scales ($k \sim 0.3 \, \hMpcinv$),
the power spectrum of RegPT damps due to the asymptotic behavior of the propagator.
On the other hand, SPT apparently works better at small scales.
However, SPT is known to accidentally give an accurate prediction around redshift $z = 1$
and is not accurate at small scales for other redshifts in general (e.g., \cite{Taruya_etal2009,Blas2014}).

In measurements from observations or simulations,
the bispectrum is usually evaluated for a given binning of wave-vectors.
To make a fair comparison with the measurements,
we need to take into account the binning effect for PT calculations.
However, the direct method is not feasible in terms of the computational cost
because there is a tremendous number of possible triangles in each bin
and it is necessary to run the calculation for each triangle.
The fast method can make it feasible with the following approximation to evaluate the binned bispectra.
The bin averaged bispectrum is given as
\begin{align}
  B_\mathrm{bin} (k_1, k_2, k_3) =& \frac{1}{V_k}
  \left[ \prod_{i=1}^{3} \int_{q_i \in [k_i - \Delta k / 2, k_i + \Delta k / 2]}
  \mathrm{d}^3 q_i \right]
  \nonumber \\
  & \times \delta_\mathrm{D} (\bm{q}_1 + \bm{q}_2 + \bm{q}_3)
  B(q_1, q_2, q_3) , \\
  V_k =& \left[ \prod_{i=1}^{3} \int_{q_i \in [k_i - \Delta k / 2, k_i + \Delta k / 2]}
  \mathrm{d}^3 q_i \right]
  \nonumber \\
  & \times \delta_\mathrm{D} (\bm{q}_1 + \bm{q}_2 + \bm{q}_3),
\end{align}
where we assume equally-spaced binning with the width $\Delta k$.
We can approximate these integrals as the summation with dense sampling ($\Delta q_i \ll q_i$):
\begin{align}
  V_k
  \simeq& 4\pi^2
  \prod_{i=1}^{3} \sum_{q_i\in [k_i - \Delta k / 2, k_i + \Delta k / 2]} q_i \Delta q_i ,
  \label{eq:binned_volume}
  \\
  B_\mathrm{bin} (k_1, k_2, k_3) \simeq& \frac{1}{V_k}
  \prod_{i=1}^{3} \sum_{q_i \in [k_i - \Delta k / 2, k_i + \Delta k / 2]}
  q_i \Delta q_i
  \nonumber \\
  & \times B(q_1, q_2, q_3) ,
  \label{eq:binned_bispectrum}
\end{align}
where the summation is carried out for linearly sampled $q_i$
in the bin range with $n_q = 50$ samples per dimension.

\begin{figure}
  \includegraphics[width=\columnwidth]{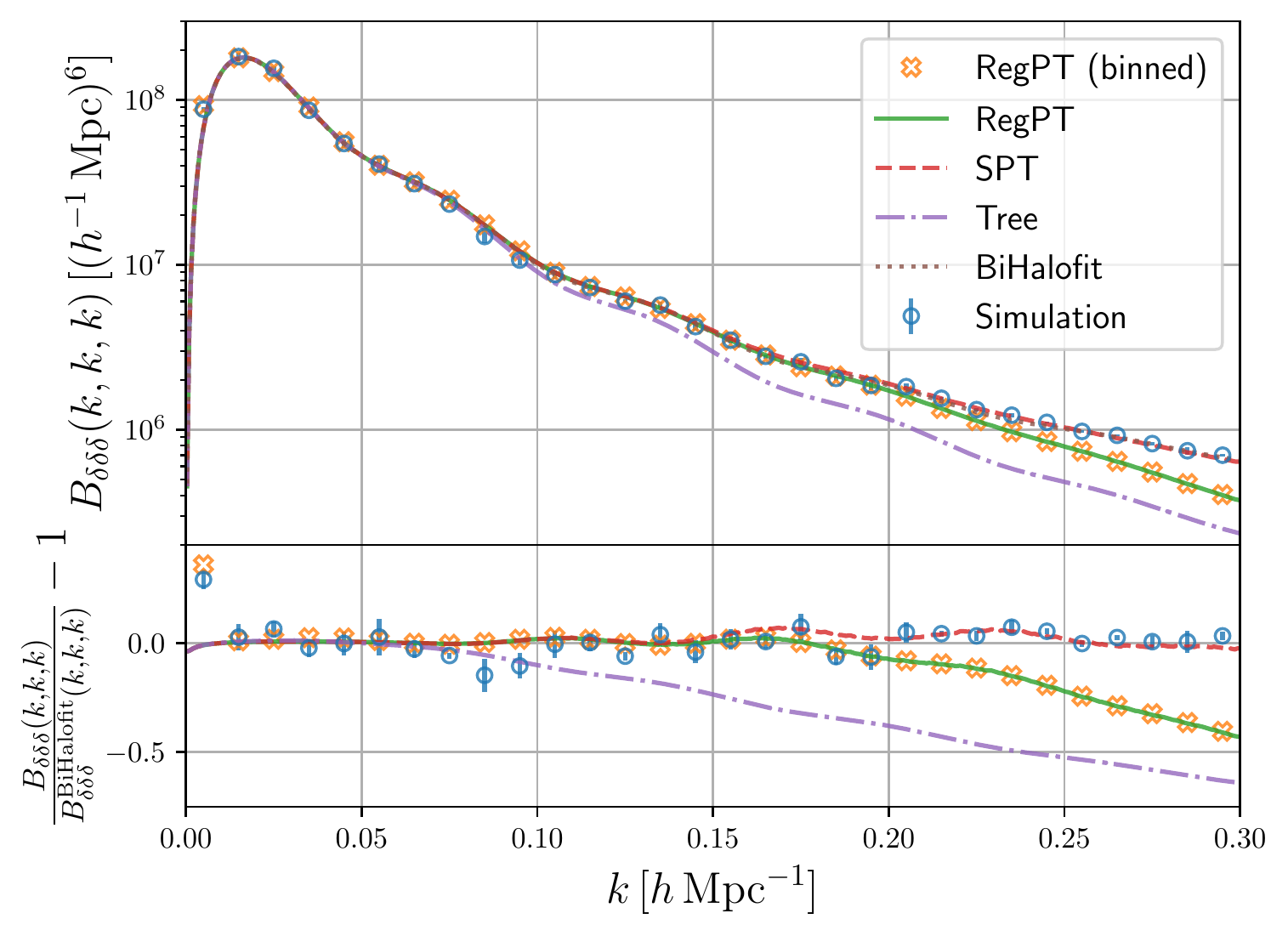}
  \caption{Comparison for density bispectra for equilateral configuration:
  measurement from $N$-body simulations,
  tree level bispectrum, unbinned bispectra based on SPT and RegPT,
  binned bispectra with the fast method, and
  the \textit{BiHalofit} fitting formula at the redshift $z = 0.901$.
  In the lower panel, the ratios in terms of \textit{BiHalofit} are shown.}
  \label{fig:Bk_equil}
\end{figure}

\begin{figure}
  \includegraphics[width=\columnwidth]{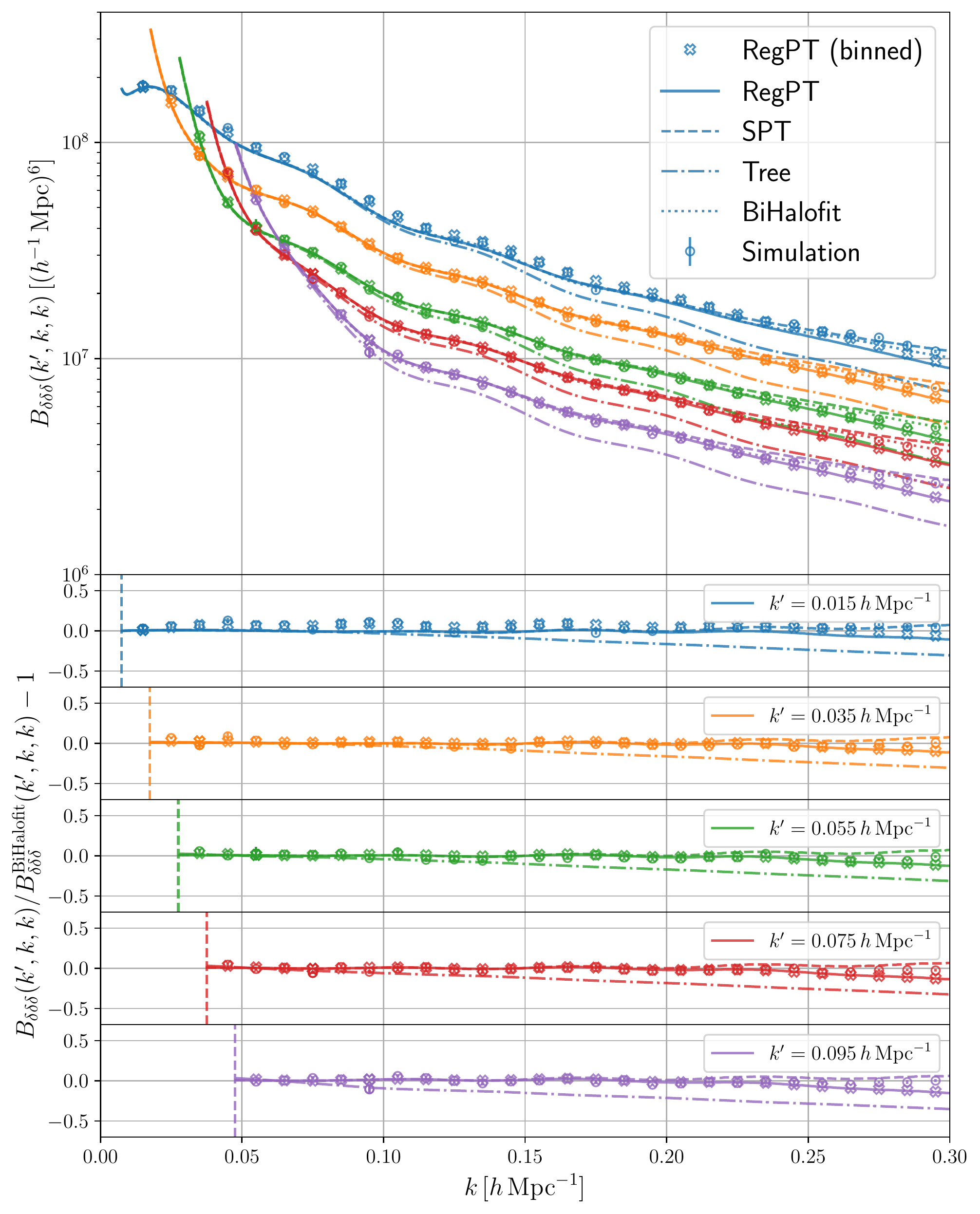}
  \caption{Comparison of density bispectra for isosceles configurations $(k', k, k)$
  with $k'/(h \, \mathrm{Mpc}^{-1}) = 0.015, 0.035, 0.055, 0.075, 0.095$:
  tree level bispectrum, unbinned bispectra based on SPT and RegPT,
  binned RegPT bispectra with the fast method, and
  the \textit{BiHalofit} fitting formula at the redshift $z = 0.901$.
  In the lower panels, the ratios in terms of \textit{BiHalofit} are shown and
  dashed lines correspond to the flattened triangle $k' = k/2$.
  }
  \label{fig:Bk_iso1}
\end{figure}

\begin{figure}
  \includegraphics[width=\columnwidth]{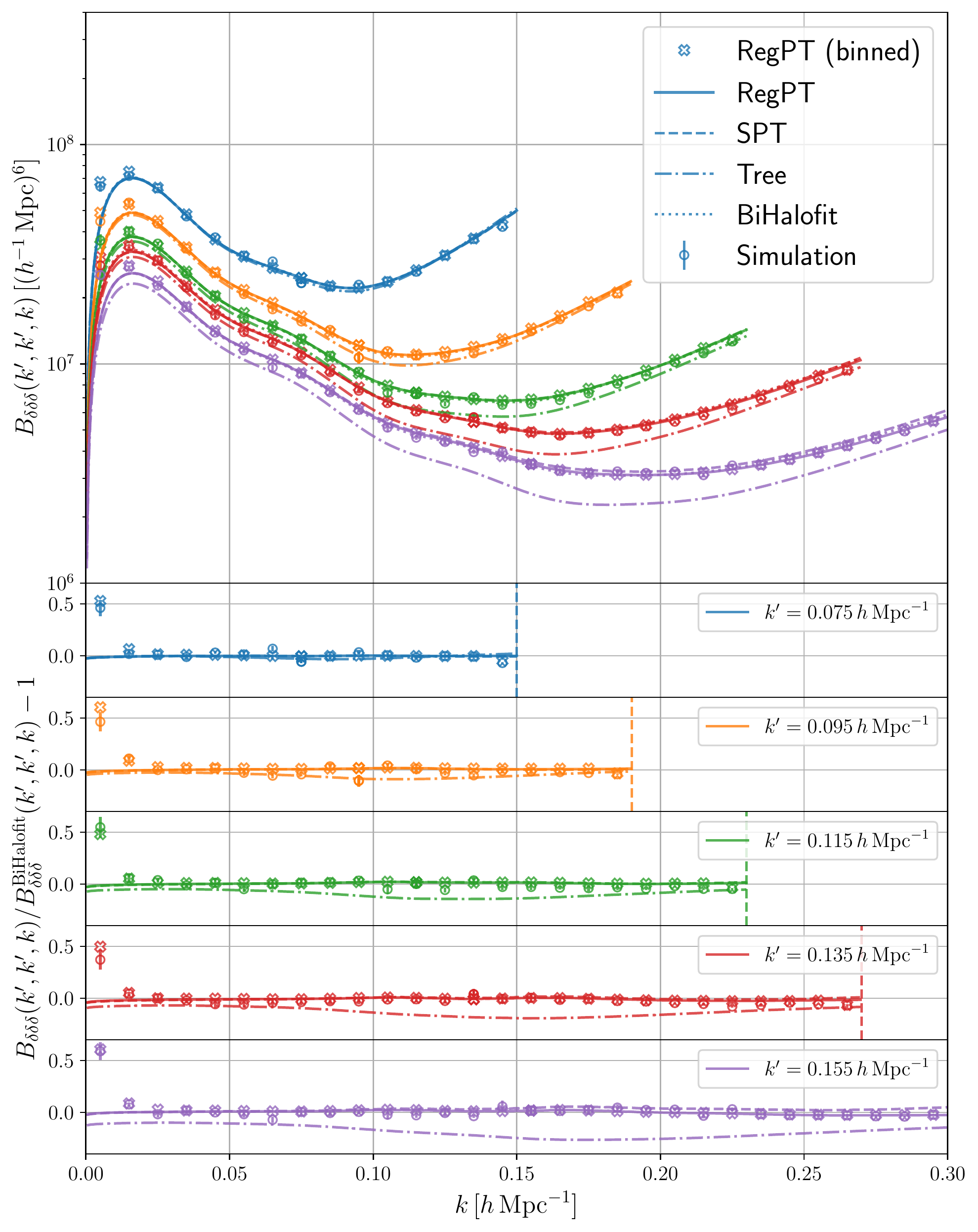}
  \caption{Comparison of density bispectra for isosceles configurations $(k', k', k)$
  with $k'/(h \, \mathrm{Mpc}^{-1}) = 0.075, 0.095, 0.115, 0.135, 0.155$:
  tree level bispectrum, unbinned bispectra based on SPT and RegPT,
  binned RegPT bispectra with the fast method, and
  the \textit{BiHalofit} fitting formula at the redshift $z = 0.901$.
  In the lower panels, the ratios in terms of \textit{BiHalofit} are shown and
  dashed lines correspond to the flattened triangle $k' = 2 k$.
  }
  \label{fig:Bk_iso2}
\end{figure}

In Figures~\ref{fig:Bk_equil}, \ref{fig:Bk_iso1}, and \ref{fig:Bk_iso2},
we show the bispectra measured from the $N$-body simulations,
the tree level calculation, the NLO PT predictions based on SPT and RegPT, and
the fitting formula \textit{BiHalofit} \cite{Takahashi2020}
for equilateral and isosceles configurations.
For comparison, we also show the binned bispectra but only with the fast method,
which are already sufficient to infer the typical impact of
the binning effect.
We have adopted the width $\Delta k = 0.01 \, \hMpcinv$.
For equilateral configurations, even at relatively large scales,
there is a sufficient number of modes and thus,
the binned and unbinned bispectra are almost identical.
On the other hand, isosceles configurations are more subject to binning effect
which persist at small scales for specific configurations and
the binned bispectra better match with the bispectra measured from the simulations.
In terms of comparison between SPT and RegPT,
up to mildly non-linear regime ($k \simeq 0.2 \, \hMpcinv$),
RegPT gives better match with the simulation results in general.
However, for equilateral configurations,
SPT performs better down to small scales ($k \simeq 0.3 \, \hMpcinv$)
but this is also just a coincidence around $z = 1$
for the same reason as in power spectrum \cite{Lazanu2018}.

\section{Conclusions}
\label{sec:conclusions}
We have developed a fast calculation method for power spectra and bispectra
in the context of the RegPT approach.
The method is based on Taylor expansions of each relevant term,
where the leading-order corrections are expressed by the
functional derivative with respect to the linear power spectrum,
effectively reducing the computational cost to one-dimension integrals.
In our computational environment \footnote{Intel Xeon E5-2695 v4 (2.1GHz)},
it takes $46$ and $112$ seconds to compute power spectra
for $120$ wave-vectors and bispectra for $10^6$ triangles,
where kernels are precomputed in our default setting,
for all combinations of density and velocity divergence, respectively.
Once these precomputations are done,
power spectra and bispectra for arbitrary wave-vectors can be
swiftly computed by interpolation of precomputed results
with adequate accuracy ($< 0.1 \%$) and
computational time for the interpolation is negligible
compared with the precomputation in a practical case.
The computation of binned bispectra requires another $47$ seconds
for linearly spaced bins with the width $\Delta k = 0.01 \, \hMpcinv$
and with $n_k = 30$ modes in each direction.
In this example, all possible bins are considered,
but if specific configurations
(e.g., only equilateral configurations) are necessary,
the calculation time becomes shorter
because linear interpolation of bispectrum requires
only data points adjacent to the configurations.
All-in-all, the calculation of the complete shape of the power spectra
and (binned) bispectra can be done in a few minutes.

We validate our method for 10 target models with 10 fiducial models sampled
by an optimized latin hypercube design, which covers a four-parameter space, and find that
the accuracy of the reconstruction is $0.5\%$ for power spectra and
$2.5\%$ for bispectra at $k = 0.3 \, \hMpcinv$
for the auto and cross spectra between the density and velocity field.
For moderate scales ($k \lesssim 0.2 \, \hMpcinv$) targeted by the current cosmological surveys,
the accuracy is expected to be better than the values quoted above.

We also compare our results with $N$-body simulations and
find that results from our fast method are consistent with simulations within a few percent level
up to mildly non-linear regime ($k \lesssim 0.2 \, \hMpcinv$).
Furthermore, we have implemented the method to incorporate the binning effect for bispectra,
which has appreciable impacts for squeezed or flattened configurations.
We find that the PT-based results agree better with the simulation results
at large scales by taking this into account,
which are more sensitive to binning effects because bispectra vary more rapidly within the bin.

The fast approach we present opens the way to conduct statistical inference
in five-dimensional fundamental cosmological parameter space rather than
with limited number of parameters (e.g., Alcock--Paczy\'nski test)
through the full shape of the power spectrum or bispectrum,
and one can investigate the constraining power of
these statistics using specific PT approaches extending the results derived in Ref.~\cite{Osato2019}.
In addition, the PT model can be applied to weak lensing analysis once nulling is
implemented \cite{Bernardeau2014} or with $k$-cut methods \cite{Taylor2018},
which allows to restrict the analysis to large scales so that the PT model prediction is applicable.
Therefore, our methods can also speed up the analysis with PT models for weak lensing.
We further note that it is possible to extend our method to combine the galaxy bias loop expansion
and explore the capability \cite{Eggemeier2020,Eggemeier2021},
since the expansion also involves high-dimensional integrals
which can be treated as functional of linear power spectrum.
And finally we stress that the fast calculation of the \textit{mixed} spectra allows the computation
of diagrams relevant for the prediction of statistics in redshift space.
In particular, the prediction of the density power spectrum in redshift space
is known to require the evaluation of the mixed bispectrum \cite{Taruya2010,Taruya2013}.
Our fast method can naturally handle this with the same efficiency.
In conclusion, the method we present potentially allows
the fast construction construct of the whole galaxy power spectrum in redshift in a few minutes.
This opens the way to the implementation of global inference methods of cosmological parameters,
in particular the five fundamental cosmological parameters.
We will address the actual implementation of such fast computation schemes
and its use in subsequent papers.

\begin{acknowledgments}
KO is supported by JSPS Overseas Research Fellowships and JSPS Research Fellowships for Young Scientists.
Numerical simulations were carried out on Cray XC50
at the Center for Computational Astrophysics.
This work was supported in part by
Grant-in-Aid for JSPS Fellows
Grant Number JP21J00011,
World Premier International Research Center Initiative (WPI Initiative),
MEXT, Japan, JSPS KAKENHI Grant Numbers JP17H06359, JP17K14273, JP19H00677, JP20H05861 and JP21H01081,
and by JST AIP Acceleration Research Grant Number JP20317829, Japan.
\end{acknowledgments}

\onecolumngrid
\appendix

\section{Expressions of power spectrum and bispectrum}
\label{sec:expressions}

\subsection{Power spectra at 2-loop level with RegPT}
First, we give formulations of power spectra at 2-loop level, i.e., up to NNLO.
The power spectra of density and velocity divergence are given as
\begin{align}
  P_{ab} (k; \eta) =& \Gamma^{(1)}_{a,\reg} (k) \Gamma^{(1)}_{b,\reg} (k) P_0 (k)+
  2 \int \frac{\mathrm{d}^3 \bm{q}}{(2 \pi)^3} \Gamma^{(2)}_{a,\reg} (\bm{q}, \bm{k}-\bm{q})
  \Gamma^{(2)}_{b,\reg} (\bm{q}, \bm{k}-\bm{q}) P_0 (q) P_0 (|\bm{k}-\bm{q}|)
  \nonumber \\
  & + 6 \int \frac{\mathrm{d}^3 \bm{q}_1}{(2 \pi)^3} \frac{\mathrm{d}^3 \bm{q}_2}{(2 \pi)^3}
  \Gamma^{(3)}_{a,\reg} (\bm{q}_1, \bm{q}_2, \bm{k}-\bm{q}_1-\bm{q}_2)
  \Gamma^{(3)}_{b,\reg} (\bm{q}_1, \bm{q}_2, \bm{k}-\bm{q}_1-\bm{q}_2)
  P_0 (q_1) P_0 (q_2) P_0 (|\bm{k}-\bm{q}_1-\bm{q}_2|) ,
\end{align}
where the regularized propagators are given as
\begin{align}
  \Gamma^{(1)}_{a,\reg} (k) &= e^\eta \left[ 1 + \alpha_k +\frac{1}{2} \alpha_k^2 +
  e^{2\eta} \bar{\Gamma}^{(1)}_{a, \text{1-loop}} (k) ( 1+\alpha_k )
  + e^{4\eta} \bar{\Gamma}^{(1)}_{a, \text{2-loop}} (k) \right] e^{-\alpha_k} , \\
  \Gamma^{(2)}_{a,\reg} (\bm{q}, \bm{k}-\bm{q}) &= e^{2\eta} \left[
  F^{(2)}_a (\bm{q}, \bm{k}-\bm{q}) (1+\alpha_k) +
  e^{2\eta} \bar{\Gamma}^{(2)}_{a, \text{1-loop}} (\bm{q}, \bm{k}-\bm{q}) \right] e^{-\alpha_k} , \\
  \Gamma^{(3)}_{a,\reg} (\bm{q}_1, \bm{q}_2, \bm{k}-\bm{q}_1-\bm{q}_2) &=
  e^{3\eta} F^{(3)}_a (\bm{q}_1, \bm{q}_2, \bm{k}-\bm{q}_1-\bm{q}_2) e^{-\alpha_k} .
\end{align}
The damping factor $\alpha_k$ is defined as
\begin{equation}
  \alpha_k \equiv \frac{k^2 \sigmad^2 e^{2\eta}}{2} ,
\end{equation}
where $\sigmad$ is defined in Eq.~\eqref{eq:sigma_d}.
The scale dependent parts of propagators are given as
\begin{align}
  \bar{\Gamma}^{(1)}_{a, \text{1-loop}} (k) =& 3 \int \frac{\mathrm{d}^3 \bm{q}}{(2 \pi)^3}
  F^{(3)}_a (\bm{q}, -\bm{q}, \bm{k}) P_0 (q) , \\
  \bar{\Gamma}^{(1)}_{a, \text{2-loop}} (k) =& 15 \int
  \frac{\mathrm{d}^3 \bm{q}_1}{(2 \pi)^3} \frac{\mathrm{d}^3 \bm{q}_2}{(2 \pi)^3}
  F^{(5)}_a (\bm{q}_1, -\bm{q}_1, \bm{q}_2, -\bm{q}_2, \bm{k})
  P_0 (q_1) P_0 (q_2) , \\
  \bar{\Gamma}^{(2)}_{a, \text{1-loop}} (\bm{k}_1, \bm{k}_2) =& 6 \int \frac{\mathrm{d}^3 \bm{q}}{(2 \pi)^3}
  F^{(4)}_a (\bm{q}, -\bm{q}, \bm{k}_1, \bm{k}_2) P_0 (q) .
\end{align}

\subsection{Bispectra at 1-loop level with RegPT}
Next, we give formulas for bispectra at 1-loop order, i.e., up to NLO.
The 1-loop bispectrum is decomposed into three parts:
\begin{equation}
  B_{abc} (\bm{k}_1, \bm{k}_2, \bm{k}_3) = B^\text{I}_{abc} (\bm{k}_1, \bm{k}_2, \bm{k}_3) +
  B^\text{II}_{abc} (\bm{k}_1, \bm{k}_2, \bm{k}_3)
  + B^\text{III}_{abc} (\bm{k}_1, \bm{k}_2, \bm{k}_3) .
\end{equation}
Each term is given as
\begin{align}
  B^\text{I}_{abc} (\bm{k}_1, \bm{k}_2, \bm{k}_3) = &
  2 \left[ \Gamma^{(2)}_{a,\reg} (\bm{k}_2, \bm{k}_3) \Gamma^{(1)}_{b,\reg} (k_2) \Gamma^{(1)}_{c,\reg} (k_3)
  P_0 (k_2) P_0 (k_3) \right. \nonumber \\
  & + \Gamma^{(2)}_{b,\reg} (\bm{k}_3, \bm{k}_1) \Gamma^{(1)}_{c,\reg} (k_3) \Gamma^{(1)}_{a,\reg} (k_1)
  P_0 (k_3) P_0 (k_1) \nonumber \\
  & \left. + \Gamma^{(2)}_{c,\reg} (\bm{k}_1, \bm{k}_2) \Gamma^{(1)}_{a,\reg} (k_1) \Gamma^{(1)}_{b,\reg} (k_2)
  P_0 (k_1) P_0 (k_2) \right] ,
\end{align}
\begin{align}
  B^\text{II}_{abc} (\bm{k}_1, \bm{k}_2, \bm{k}_3) = & 8 \int \frac{d^3 \bm{q}}{(2\pi)^3}
  \Gamma^{(2)}_{a,\reg} (\bm{k}_1-\bm{q}, \bm{q}) \Gamma^{(2)}_{b,\reg} (\bm{k}_2+\bm{q}, -\bm{q})
  \Gamma^{(2)}_{c,\reg} (-\bm{q}-\bm{k}_2, -\bm{k}_1+\bm{q}) \nonumber \\
  & \times P_0 (|\bm{k}_1 - \bm{q}|) P_0 (|\bm{k}_2 + \bm{q}|) P_0 (q) ,
\end{align}
\begin{align}
  B^\text{III}_{abc} (\bm{k}_1, \bm{k}_2, \bm{k}_3) = & 6 \int \frac{\mathrm{d}^3 \bm{q}}{(2\pi)^3} \times \nonumber \\
  & \left[ \Gamma^{(3)}_{a,\reg} (-\bm{k}_3, -\bm{k}_2+\bm{q}, -\bm{q})
  \Gamma^{(2)}_{b,\reg} (\bm{k}_2-\bm{q}, \bm{q})
  \Gamma^{(1)}_{c,\reg} (k_3) P_0 (|\bm{k}_2-\bm{q}|) P_0 (k_3) P_0 (q) \right. \nonumber \\
  & + \Gamma^{(3)}_{a,\reg} (-\bm{k}_2, -\bm{k}_3+\bm{q}, -\bm{q})
  \Gamma^{(2)}_{c,\reg} (\bm{k}_3-\bm{q}, \bm{q})
  \Gamma^{(1)}_{b,\reg} (k_2) P_0 (|\bm{k}_3-\bm{q}|) P_0 (k_2) P_0 (q) \nonumber \\
  & + \Gamma^{(3)}_{b,\reg} (-\bm{q}, -\bm{k}_1+\bm{q}, -\bm{k}_3)
  \Gamma^{(2)}_{a,\reg} (\bm{k}_1-\bm{q}, \bm{q})
  \Gamma^{(1)}_{c,\reg} (k_3) P_0 (|\bm{k}_1-\bm{q}|) P_0 (k_3) P_0 (q) \nonumber \\
  & + \Gamma^{(3)}_{b,\reg} (-\bm{k}_1, -\bm{k}_3+\bm{q}, -\bm{q})
  \Gamma^{(2)}_{c,\reg} (\bm{k}_3-\bm{q}, \bm{q})
  \Gamma^{(1)}_{a,\reg} (k_1) P_0 (|\bm{k}_3-\bm{q}|) P_0 (k_1) P_0 (q) \nonumber \\
  & + \Gamma^{(3)}_{c,\reg} (-\bm{k}_1+\bm{q}, -\bm{q}, -\bm{k}_2)
  \Gamma^{(2)}_{a,\reg} (\bm{k}_1-\bm{q}, \bm{q})
  \Gamma^{(1)}_{b,\reg} (k_2) P_0 (|\bm{k}_1-\bm{q}|) P_0 (k_2) P_0 (q) \nonumber \\
  & \left. + \Gamma^{(3)}_{c,\reg} (-\bm{k}_1, -\bm{q}, -\bm{k}_2+\bm{q})
  \Gamma^{(2)}_{b,\reg} (\bm{k}_2-\bm{q}, \bm{q})
  \Gamma^{(1)}_{a,\reg} (k_1) P_0 (|\bm{k}_2-\bm{q}|) P_0 (k_1) P_0 (q) \right] .
\end{align}
In order to keep the order of each part as 2-loop level,
the regularized propagators are computed at 1-loop level for $B^\text{I}_{abc}$
and tree level for $B^\text{II}_{abc}$ and $B^\text{III}_{abc}$.
The 1-loop level propagators are given as
\begin{align}
  \Gamma^{(1)}_{a,\reg} (k) =& e^\eta \left[ 1 + \alpha_k +
  e^{2\eta} \bar{\Gamma}^{(1)}_{a, \text{1-loop}} (k) \right] e^{-\alpha_k} , \\
  \Gamma^{(2)}_{a,\reg} (\bm{k}_1, \bm{k}_2) =& e^{2\eta} \left[
  F^{(2)}_a (\bm{k}_1, \bm{k}_2) (1+\alpha_k)
  + e^{2\eta} \bar{\Gamma}^{(2)}_{a, \text{1-loop}} (\bm{k}_1, \bm{k}_2) \right]
  e^{-\alpha_k} ,
\end{align}
and the tree level propagators are given as
\begin{equation}
\Gamma^{(n)}_{a,\reg} (\bm{k}_1, \bm{k}_2, \ldots, \bm{k}_n) = e^{n \eta}
F^{(n)}_a (\bm{k}_1, \bm{k}_2, \ldots, \bm{k}_n) e^{-\alpha_k} ,
\end{equation}
where $k = |\bm{k}_1 + \bm{k}_2 + \cdots + \bm{k}_n |$.

\subsection{IR-safe integrand}
In the loop integrals at large scales
IR cancellation occurs and this effect might lead to numerically unstable results.
Here, we give IR-safe integrand for 1-loop bispectra based on Ref.~\cite{Baldauf2015},
where IR cancellation is taken into account beforehand.
First, we define the integrands of bispectra:
\begin{equation}
B^{\text{II}}_{abc} (\bm{k}_1, \bm{k}_2, \bm{k}_3) \equiv
\int \frac{\mathrm{d}^3 \bm{q}}{(2 \pi)^3} b^{\text{II}}_{abc} (\bm{q}, \bm{k}_1, \bm{k}_2), \
B^{\text{III}}_{abc} (\bm{k}_1, \bm{k}_2, \bm{k}_3) \equiv
\int \frac{\mathrm{d}^3 \bm{q}}{(2 \pi)^3} \left[ b^{\text{III}}_{abc} (\bm{q}, \bm{k}_2, \bm{k}_3)
+ \text{5 perms.} \right].
\end{equation}
Then, the integrands of IR-safe terms are given by
\begin{align}
\tilde{b}^{\text{II}}_{abc} (\bm{q}, \bm{k}_1, \bm{k}_2, \bm{k}_3) =&
\frac{1}{2} \left\{ \left[
b^{\text{II}}_{abc} (\bm{q}, \bm{k}_1, \bm{k}_2) \Theta (r_1-q) \Theta (p_2-q) +
b^{\text{II}}_{abc} (-\bm{q}, \bm{k}_1, \bm{k}_2) \Theta (p_1-q) \Theta (r_2-q) \right]
\right. \nonumber \\
& \left. + [\text{perm.}: \bm{k}_1 \to \bm{k}_3]
+ [\text{perm.}: \bm{k}_2 \to \bm{k}_3]
\right\} , \\
\tilde{b}^{\text{III}}_{abc} (\bm{q}, \bm{k}_1, \bm{k}_2, \bm{k}_3) =&
\left[ b^{\text{III}}_{abc} (\bm{q}, \bm{k}_2, \bm{k}_3) \Theta (p_2-q) +
b^{\text{III}}_{abc} (-\bm{q}, \bm{k}_2, \bm{k}_3) \Theta (r_2-q) \right] +
\text{5 perms.} ,
\end{align}
where $\Theta$ is the step function and we introduce following notations:
\begin{equation}
\bm{p}_i \equiv \bm{k}_i - \bm{q}, \ \bm{r}_i \equiv \bm{k}_i + \bm{q}
\ (i = 1, 2, 3).
\end{equation}
Note that the $B^{\text{I}}_{abc}$ term is not subject to IR cancellation.

Then, the IR-safe forms are given as
\begin{align}
B^\text{II}_{abc} (\bm{k}_1, \bm{k}_2, \bm{k}_3) =& 8 \int \frac{\mathrm{d}^3 \bm{q}}{(2\pi)^3} \times
\nonumber \\
& \left[ \Theta (p_1 - q) \Theta (r_2 - q)
\Gamma^{(2)}_{a,\reg} (\bm{p}_1, \bm{q}) \Gamma^{(2)}_{b,\reg} (\bm{r}_2, -\bm{q})
\Gamma^{(2)}_{c,\reg} (-\bm{r}_2, -\bm{p}_1) P_0 (p_1) P_0 (r_2) P_0 (q)
\right. \nonumber \\
& + \Theta (r_1 - q) \Theta (p_2 - q)
\Gamma^{(2)}_{a,\reg} (\bm{r}_1, -\bm{q}) \Gamma^{(2)}_{b,\reg} (\bm{p}_2, \bm{q})
\Gamma^{(2)}_{c,\reg} (-\bm{p}_2, -\bm{r}_1) P_0 (r_1) P_0 (p_2) P_0 (q)
\nonumber \\
& + \Theta (p_3 - q) \Theta (r_2 - q)
\Gamma^{(2)}_{c,\reg} (\bm{p}_3, \bm{q}) \Gamma^{(2)}_{b,\reg} (\bm{r}_2, -\bm{q})
\Gamma^{(2)}_{a,\reg} (-\bm{r}_2, -\bm{p}_3) P_0 (p_3) P_0 (r_2) P_0 (q)
\nonumber \\
& + \Theta (r_3 - q) \Theta (p_2 - q)
\Gamma^{(2)}_{c,\reg} (\bm{r}_3, -\bm{q}) \Gamma^{(2)}_{b,\reg} (\bm{p}_2, \bm{q})
\Gamma^{(2)}_{a,\reg} (-\bm{p}_2, -\bm{r}_3) P_0 (r_3) P_0 (p_2) P_0 (q)
\nonumber \\
& + \Theta (p_1 - q) \Theta (r_3 - q)
\Gamma^{(2)}_{a,\reg} (\bm{p}_1, \bm{q}) \Gamma^{(2)}_{c,\reg} (\bm{r}_3, -\bm{q})
\Gamma^{(2)}_{b,\reg} (-\bm{r}_3, -\bm{p}_1) P_0 (p_1) P_0 (r_3) P_0 (q)
\nonumber \\
& \left. + \Theta (r_1 - q) \Theta (p_3 - q)
\Gamma^{(2)}_{a,\reg} (\bm{r}_1, -\bm{q}) \Gamma^{(2)}_{c,\reg} (\bm{p}_3, \bm{q})
\Gamma^{(2)}_{b,\reg} (-\bm{p}_3, -\bm{r}_1) P_0 (r_1) P_0 (p_3) P_0 (q)
\right] ,
\end{align}
\begin{align}
B^\text{III}_{abc} (\bm{k}_1, \bm{k}_2, \bm{k}_3) =& 6 \int \frac{\mathrm{d}^3 \bm{q}}{(2\pi)^3} \times
\nonumber \\
& \left\{ \Theta (p_2 - q) \left[
\Gamma^{(3)}_{a,\reg} (-\bm{k}_3, -\bm{p}_2, -\bm{q})
\Gamma^{(2)}_{b,\reg} (\bm{p}_2, \bm{q})
\Gamma^{(1)}_{c,\reg} (k_3) P_0 (p_2) P_0 (k_3) P_0 (q)
\right. \right. \nonumber \\
& \left. +
\Gamma^{(3)}_{c,\reg} (-\bm{k}_1, -\bm{q}, -\bm{p}_2)
\Gamma^{(2)}_{b,\reg} (\bm{p}_2, \bm{q})
\Gamma^{(1)}_{a,\reg} (k_1) P_0 (p_2) P_0 (k_1) P_0 (q) \right]
\nonumber \\
& + \Theta (r_2 - q) \left[
\Gamma^{(3)}_{a,\reg} (-\bm{k}_3, -\bm{r}_2, -\bm{q})
\Gamma^{(2)}_{b,\reg} (\bm{r}_2, \bm{q})
\Gamma^{(1)}_{c,\reg} (k_3) P_0 (r_2) P_0 (k_3) P_0 (q)
\right. \nonumber \\
& \left. +
\Gamma^{(3)}_{c,\reg} (-\bm{k}_1, -\bm{q}, -\bm{r}_2)
\Gamma^{(2)}_{b,\reg} (\bm{r}_2, \bm{q})
\Gamma^{(1)}_{a,\reg} (k_1) P_0 (r_2) P_0 (k_1) P_0 (q) \right]
\nonumber \\
& + \Theta (p_3 - q) \left[
\Gamma^{(3)}_{a,\reg} (-\bm{k}_2, -\bm{p}_3, -\bm{q})
\Gamma^{(2)}_{c,\reg} (\bm{p}_3, \bm{q})
\Gamma^{(1)}_{b,\reg} (k_2) P_0 (p_3) P_0 (k_2) P_0 (q)
\right. \nonumber \\
& \left. +
\Gamma^{(3)}_{b,\reg} (-\bm{k}_1, -\bm{p}_3, -\bm{q})
\Gamma^{(2)}_{c,\reg} (\bm{p}_3, \bm{q})
\Gamma^{(1)}_{a,\reg} (k_1) P_0 (p_3) P_0 (k_1) P_0 (q) \right]
\nonumber \\
& + \Theta (r_3 - q) \left[
\Gamma^{(3)}_{a,\reg} (-\bm{k}_2, -\bm{r}_3, -\bm{q})
\Gamma^{(2)}_{c,\reg} (\bm{r}_3, \bm{q})
\Gamma^{(1)}_{b,\reg} (k_2) P_0 (r_3) P_0 (k_2) P_0 (q)
\right. \nonumber \\
& \left. +
\Gamma^{(3)}_{b,\reg} (-\bm{k}_1, -\bm{r}_3, -\bm{q})
\Gamma^{(2)}_{c,\reg} (\bm{r}_3, \bm{q})
\Gamma^{(1)}_{a,\reg} (k_1) P_0 (r_3) P_0 (k_1) P_0 (q) \right]
\nonumber \\
& + \Theta (p_1 - q) \left[
\Gamma^{(3)}_{b,\reg} (-\bm{q}, -\bm{p}_1, -\bm{k}_3)
\Gamma^{(2)}_{a,\reg} (\bm{p}_1, \bm{q})
\Gamma^{(1)}_{c,\reg} (k_3) P_0 (p_1) P_0 (k_3) P_0 (q)
\right. \nonumber \\
& \left. +
\Gamma^{(3)}_{c,\reg} (-\bm{p}_1, -\bm{q}, -\bm{k}_2)
\Gamma^{(2)}_{a,\reg} (\bm{p}_1, \bm{q})
\Gamma^{(1)}_{b,\reg} (k_2) P_0 (p_1) P_0 (k_2) P_0 (q) \right]
\nonumber \\
& + \Theta (r_1 - q) \left[
\Gamma^{(3)}_{b,\reg} (-\bm{q}, -\bm{r}_1, -\bm{k}_3)
\Gamma^{(2)}_{a,\reg} (\bm{r}_1, \bm{q})
\Gamma^{(1)}_{c,\reg} (k_3) P_0 (r_1) P_0 (k_3) P_0 (q)
\right. \nonumber \\
& \left. \left. +
\Gamma^{(3)}_{c,\reg} (-\bm{r}_1, -\bm{q}, -\bm{k}_2)
\Gamma^{(2)}_{a,\reg} (\bm{r}_1, \bm{q})
\Gamma^{(1)}_{b,\reg} (k_2) P_0 (r_1) P_0 (k_2) P_0 (q) \right]
\right\} .
\end{align}
For the correction terms, we do not incorporate IR-safe integrand
because the IR divergence of the corrections is already small.

\subsection{Expressions with SPT}
\label{sec:expressions_SPT}
It is also possible to construct expressions of power spectrum and bispectrum based on SPT with propagators.
We keep the terms up to the third order of linear power spectrum of RegPT expressions
for power spectrum at 2-loop level and bispectrum at 1-loop level.

For SPT power spectra at 2-loop level, the expressions are given as
\begin{align}
  P^\text{SPT}_{ab} (k ; \eta) =& e^{2\eta} P_0 (k) + e^{4 \eta} \left( \bar{\Gamma}^{(1)}_{a, \text{1-loop}} (k) +
  \bar{\Gamma}^{(1)}_{b, \text{1-loop}} (k) \right) P_0 (k)
  \nonumber \\
  & + 2 e^{4\eta} \int \frac{\mathrm{d}^3 \bm{q}}{(2 \pi)^3}
  F^{(2)}_a (\bm{q}, \bm{k}-\bm{q}) F^{(2)}_b (\bm{q}, \bm{k}-\bm{q}) P_0 (q) P_0 (|\bm{k}-\bm{q}|)
  \nonumber \\
  & + e^{6\eta} \left( \bar{\Gamma}^{(1)}_{a, \text{1-loop}} (k) \bar{\Gamma}^{(1)}_{b, \text{1-loop}} (k) +
  \bar{\Gamma}^{(1)}_{a, \text{2-loop}} (k) + \bar{\Gamma}^{(1)}_{b, \text{2-loop}} (k) \right) P_0 (k)
  \nonumber \\
  & + 2 e^{6\eta} \int \frac{\mathrm{d}^3 \bm{q}}{(2 \pi)^3}
  \left[ F^{(2)}_a (\bm{q}, \bm{k}-\bm{q}) \bar{\Gamma}^{(2)}_{b, \text{1-loop}} (\bm{q}, \bm{k}-\bm{q}) +
  \bar{\Gamma}^{(2)}_{a, \text{1-loop}} (\bm{q}, \bm{k}-\bm{q}) F^{(2)}_b (\bm{q}, \bm{k}-\bm{q})  \right] P_0 (q) P_0 (|\bm{k}-\bm{q}|)
  \nonumber \\
  & + 6 e^{6\eta} \int \frac{\mathrm{d}^3 \bm{q}_1}{(2 \pi)^3} \frac{\mathrm{d}^3 \bm{q}_2}{(2 \pi)^3}
  F^{(3)}_a (\bm{q}_1, \bm{q}_2, \bm{k}-\bm{q}_1-\bm{q}_2)
  F^{(3)}_b (\bm{q}_1, \bm{q}_2, \bm{k}-\bm{q}_1-\bm{q}_2)
  P_0 (q_1) P_0 (q_2) P_0 (|\bm{k}-\bm{q}_1-\bm{q}_2|) .
\end{align}

For SPT bispectrum at 1-loop level, the bispectrum is decomposed into three terms:
\begin{equation}
  B^\text{SPT}_{abc} (\bm{k}_1, \bm{k}_2, \bm{k}_3) = B^\text{SPT,I}_{abc} (\bm{k}_1, \bm{k}_2, \bm{k}_3) +
  B^\text{SPT,II}_{abc} (\bm{k}_1, \bm{k}_2, \bm{k}_3)
  + B^\text{SPT,III}_{abc} (\bm{k}_1, \bm{k}_2, \bm{k}_3) ,
\end{equation}
and each term is given as
\begin{align}
  B^\text{SPT,I}_{abc} (\bm{k}_1, \bm{k}_2, \bm{k}_3) = &
  2 e^{4 \eta} \left[
  F^{(2)}_a (\bm{k}_2, \bm{k}_3) P_0 (k_2) P_0 (k_3) +
  F^{(2)}_b (\bm{k}_3, \bm{k}_1) P_0 (k_3) P_0 (k_1) +
  F^{(2)}_c (\bm{k}_1, \bm{k}_2) P_0 (k_1) P_0 (k_2) \right]
  \nonumber \\
  & + 2 e^{6 \eta} \left[ \bar{\Gamma}^{(2)}_{a, \text{1-loop}} (\bm{k}_2, \bm{k}_3) + F^{(2)}_a (\bm{k}_2, \bm{k}_3)
  \left( \bar{\Gamma}^{(1)}_{b, \text{1-loop}} (k_2) + \bar{\Gamma}^{(1)}_{c, \text{1-loop}} (k_3) \right)
  \right] P_0 (k_2) P_0 (k_3)
  \nonumber \\
  & + 2 e^{6 \eta} \left[ \bar{\Gamma}^{(2)}_{b, \text{1-loop}} (\bm{k}_3, \bm{k}_1) + F^{(2)}_b (\bm{k}_3, \bm{k}_1)
  \left( \bar{\Gamma}^{(1)}_{c, \text{1-loop}} (k_3) + \bar{\Gamma}^{(1)}_{a, \text{1-loop}} (k_1) \right)
  \right] P_0 (k_3) P_0 (k_1)
  \nonumber \\
  & + 2 e^{6 \eta} \left[ \bar{\Gamma}^{(2)}_{c, \text{1-loop}} (\bm{k}_1, \bm{k}_2) + F^{(2)}_c (\bm{k}_1, \bm{k}_2)
  \left( \bar{\Gamma}^{(1)}_{a, \text{1-loop}} (k_1) + \bar{\Gamma}^{(1)}_{b, \text{1-loop}} (k_2) \right)
  \right] P_0 (k_1) P_0 (k_2) ,
\end{align}
\begin{align}
  B^\text{SPT,II}_{abc} (\bm{k}_1, \bm{k}_2, \bm{k}_3) = & 8 e^{6\eta} \int \frac{\mathrm{d}^3 \bm{q}}{(2\pi)^3}
  F^{(2)}_a (\bm{k}_1-\bm{q}, \bm{q}) F^{(2)}_b (\bm{k}_2+\bm{q}, -\bm{q})
  F^{(2)}_c (-\bm{q}-\bm{k}_2, -\bm{k}_1+\bm{q}) \nonumber \\
  & \times P_0 (|\bm{k}_1 - \bm{q}|) P_0 (|\bm{k}_2 + \bm{q}|) P_0 (q) ,
\end{align}
\begin{align}
  B^\text{SPT,III}_{abc} (\bm{k}_1, \bm{k}_2, \bm{k}_3) = & 6 e^{6\eta} \int \frac{\mathrm{d}^3 \bm{q}}{(2\pi)^3} \times \nonumber \\
  & \left[ F^{(3)}_a (-\bm{k}_3, -\bm{k}_2+\bm{q}, -\bm{q})
  F^{(2)}_b (\bm{k}_2-\bm{q}, \bm{q})
  P_0 (|\bm{k}_2-\bm{q}|) P_0 (k_3) P_0 (q) \right. \nonumber \\
  & + F^{(3)}_a (-\bm{k}_2, -\bm{k}_3+\bm{q}, -\bm{q})
  F^{(2)}_c (\bm{k}_3-\bm{q}, \bm{q})
  P_0 (|\bm{k}_3-\bm{q}|) P_0 (k_2) P_0 (q) \nonumber \\
  & + F^{(3)}_b (-\bm{q}, -\bm{k}_1+\bm{q}, -\bm{k}_3)
  F^{(2)}_a (\bm{k}_1-\bm{q}, \bm{q})
  P_0 (|\bm{k}_1-\bm{q}|) P_0 (k_3) P_0 (q) \nonumber \\
  & + F^{(3)}_b (-\bm{k}_1, -\bm{k}_3+\bm{q}, -\bm{q})
  F^{(2)}_c (\bm{k}_3-\bm{q}, \bm{q})
  P_0 (|\bm{k}_3-\bm{q}|) P_0 (k_1) P_0 (q) \nonumber \\
  & + F^{(3)}_c (-\bm{k}_1+\bm{q}, -\bm{q}, -\bm{k}_2)
  F^{(2)}_a (\bm{k}_1-\bm{q}, \bm{q})
  P_0 (|\bm{k}_1-\bm{q}|) P_0 (k_2) P_0 (q) \nonumber \\
  &\left. + F^{(3)}_c (-\bm{k}_1, -\bm{q}, -\bm{k}_2+\bm{q})
  F^{(2)}_b (\bm{k}_2-\bm{q}, \bm{q})
  P_0 (|\bm{k}_2-\bm{q}|) P_0 (k_1) P_0 (q) \right] .
\end{align}

\section{Expressions for kernels}
\label{sec:kernels}

\subsection{Kernels for power spectra}
Here, we give expressions of correction terms with response function approach for power spectra (Eq.~\ref{eq:rec_pk}).
The correction terms for power spectra at the leading order are given as
\begin{align}
\delta P_{ab} (k; \eta) =& \delta \Gamma^{(1)}_{a,\reg} (k) \Gamma^{(1)}_{b,\reg} (k) \Pfid_0 (k)+
\Gamma^{(1)}_{a,\reg} (k) \delta \Gamma^{(1)}_{b,\reg} (k) \Pfid_0 (k) +
\Gamma^{(1)}_{a,\reg} (k) \Gamma^{(1)}_{b,\reg} (k) \delta P_0 (k) \nonumber \\
& + 2 \int \frac{\mathrm{d}^3 \bm{q}}{(2 \pi)^3} \left[ 2 \Gamma^{(2)}_{a,\reg} (\bm{q}, \bm{k}-\bm{q}) \Gamma^{(2)}_{b,\reg} (\bm{q}, \bm{k}-\bm{q})
\Pfid_0 (|\bm{k}-\bm{q}|) \delta P_0 (q) \right. \nonumber \\
& + \delta \Gamma^{(2)}_{a,\reg} (\bm{q}, \bm{k}-\bm{q})
\Gamma^{(2)}_{b,\reg} (\bm{q}, \bm{k}-\bm{q})
\Pfid_0 (q) \Pfid_0 (|\bm{k}-\bm{q}|) \nonumber \\
& + \left. \Gamma^{(2)}_{a,\reg} (\bm{q}, \bm{k}-\bm{q})
\delta \Gamma^{(2)}_{b,\reg} (\bm{q}, \bm{k}-\bm{q})
\Pfid_0 (q) \Pfid_0 (|\bm{k}-\bm{q}|) \right] \nonumber \\
& + 18 \int \frac{\mathrm{d}^3 \bm{q}_1}{(2 \pi)^3} \frac{\mathrm{d}^3 \bm{q}_2}{(2 \pi)^3}
\Gamma^{(3)}_{a,\reg} (\bm{q}_1, \bm{q}_2, \bm{k}-\bm{q}_1-\bm{q}_2)
\Gamma^{(3)}_{b,\reg} (\bm{q}_1, \bm{q}_2, \bm{k}-\bm{q}_1-\bm{q}_2)
\Pfid_0 (q_2) \Pfid_0 (|\bm{k}-\bm{q}_1-\bm{q}_2|) \delta P_0 (q_1) .
\nonumber \\
\end{align}
Note that $\Gamma^{(3)}_{a,\reg}$ does not involve linear power spectrum at tree level
and thus $\delta \Gamma^{(3)}_{a,\reg} = 0$.
The first-order corrections for propagators are
\begin{align}
\delta \Gamma^{(1)}_{a,\reg} (k) &= e^{3\eta}
\left[ \delta \bar{\Gamma}^{(1)}_{a, \text{1-loop}} (k) ( 1+\alpha_k )
+ e^{2\eta} \delta \bar{\Gamma}^{(1)}_{a, \text{2-loop}} (k) \right] e^{-\alpha_k} , \\
\delta \Gamma^{(2)}_{a,\reg} (\bm{q}, \bm{k}-\bm{q}) &= e^{4\eta}
\delta \bar{\Gamma}^{(2)}_{a, \text{1-loop}} (\bm{q}, \bm{k}-\bm{q}) e^{-\alpha_k} ,
\end{align}
where scale-dependent parts are given as
\begin{align}
\delta \bar{\Gamma}^{(1)}_{a, \text{1-loop}} (k) &= 3 \int \frac{\mathrm{d}^3 \bm{q}}{(2 \pi)^3}
F^{(3)}_a (\bm{q}, -\bm{q}, \bm{k}) \delta P_0 (q) , \\
\delta \bar{\Gamma}^{(1)}_{a, \text{2-loop}} (k) &= 30 \int
\frac{\mathrm{d}^3 \bm{q}_1}{(2 \pi)^3} \frac{\mathrm{d}^3 \bm{q}_2}{(2 \pi)^3}
F^{(5)}_a (\bm{q}_1, -\bm{q}_1, \bm{q}_2, -\bm{q}_2, \bm{k}) \Pfid_0 (q_1) \delta P_0 (q_2) , \\
\delta \bar{\Gamma}^{(2)}_{a, \text{1-loop}} (\bm{k}_1, \bm{k}_2) &= 6 \int \frac{\mathrm{d}^3 \bm{q}}{(2 \pi)^3}
F^{(4)}_a (\bm{q}, -\bm{q}, \bm{k}_1, \bm{k}_2) \delta P_0 (q) .
\end{align}

Next, we rewrite the expressions in the form of one-dimensional integration.
First, we define angular averaged kernels as,
\begin{align}
I_a (q; k) &= 3 \int \frac{\mathrm{d}^2 \bm{\Omega}_q}{4 \pi}
F^{(3)}_a (\bm{q}, -\bm{q}, \bm{k}) , \\
J_a (q_1, q_2; k) &= 15 \int \frac{\mathrm{d}^2 \bm{\Omega}_{q_1}}{4 \pi} \frac{\mathrm{d}^2 \bm{\Omega}_{q_2}}{4 \pi}
F^{(5)}_a (\bm{q}_1, -\bm{q}_1, \bm{q}_2, -\bm{q}_2, \bm{k}), \\
K_a (q; k_1, k_2, k_3) &= 6 \int \frac{\mathrm{d}^2 \bm{\Omega}_q}{4 \pi}
F^{(4)}_a (\bm{q}, -\bm{q}, \bm{k}_1, \bm{k}_2) ,
\end{align}
where $\bm{k}_3 = \bm{k}_1 + \bm{k}_2$.
Note that these functions are independent of cosmology.
We can compute these functions in advance to speed up
calculations.
Then, we can rewrite the scale-dependent parts of the propagators as
\begin{align}
\delta \bar{\Gamma}^{(1)}_{a, \text{1-loop}} (k) &= \int \frac{q^2 \mathrm{d} q}{2 \pi^2} L^{(1)}_a (q; k) \delta P_0 (q), \\
\delta \bar{\Gamma}^{(1)}_{a, \text{2-loop}} (k) &= 2 \int \frac{q^2 \mathrm{d} q}{2 \pi^2} M^{(1)}_a (q; k) \delta P_0 (q), \\
\delta \bar{\Gamma}^{(2)}_{a, \text{1-loop}} (\bm{k}_1, \bm{k}_2) &= \int \frac{q^2 \mathrm{d} q}{2 \pi^2}
N^{(2)}_a (q; k_1, k_2, k_3) \delta P_0 (q) ,
\end{align}
where
\begin{align}
L^{(1)}_a (q; k) &= I_a (q; k), \\
M^{(1)}_a (q; k) &= \int \frac{p^2 \mathrm{d} p}{2 \pi^2} J_a (p, q; k) \Pfid_0 (p) , \\
N^{(2)}_a (q; k_1, k_2, k_3) &= K_a (q; k_1, k_2, k_3) .
\end{align}

Next, we rewrite the whole expression of the 2-loop spectra:
\begin{align}
& \int \frac{\mathrm{d}^3 \bm{q}}{(2 \pi)^3} \Gamma^{(2)}_{a,\reg} (\bm{q}, \bm{k}-\bm{q}; \eta) \Gamma^{(2)}_{b,\reg} (\bm{q}, \bm{k}-\bm{q}; \eta)
\Pfid_0 (|\bm{k}-\bm{q}|) \delta P_0 (q) \nonumber \\
=& e^{4 \eta} e^{-2 \alpha_k} \left[ (1+\alpha_k)^2 \int \frac{q^2 \mathrm{d}q}{2 \pi^2}
X^{(2)}_{ab} (q, k) \delta P_0 (q) + e^{2\eta} (1+\alpha_k)  \int \frac{q^2 \mathrm{d}q}{2 \pi^2}
\left( Y^{(2)}_{ab} (q, k) + Y^{(2)}_{ba} (q, k) \right) \delta P_0 (q) \right. \nonumber \\
&+ \left. e^{4\eta} \int \frac{q^2 \mathrm{d}q}{2 \pi^2} Z^{(2)}_{ab} (q, k) \delta P_0 (q) \right] \\
& \int \frac{\mathrm{d}^3 \bm{q}}{(2 \pi)^3} \Gamma^{(2)}_{a,\reg} (\bm{q}, \bm{k}-\bm{q}; \eta)
\delta \Gamma^{(2)}_{b,\reg} (\bm{q}, \bm{k}-\bm{q}; \eta) \Pfid_0 (q) \Pfid_0 (|\bm{k}-\bm{q}|) \nonumber \\
=& e^{6\eta} e^{-2 \alpha_k} \left[ (1+\alpha_k) \int \frac{q^2 \mathrm{d}q}{2 \pi^2}
Q^{(2)}_{ab} (q, k) \delta P_0 (q)
+ e^{2 \eta} \int \frac{q^2 \mathrm{d}q}{2 \pi^2} R^{(2)}_{ab} (q, k) \delta P_0 (q) \right] , \\
& \int \frac{\mathrm{d}^3 \bm{q}_1}{(2 \pi)^3} \frac{\mathrm{d}^3 \bm{q}_2}{(2 \pi)^3}
\Gamma^{(3)}_{a,\reg} (\bm{q}_1, \bm{q}_2, \bm{k}-\bm{q}_1-\bm{q}_2; \eta)
\Gamma^{(3)}_{b,\reg} (\bm{q}_1, \bm{q}_2, \bm{k}-\bm{q}_1-\bm{q}_2; \eta)
\Pfid_0 (q_2) \Pfid_0 (|\bm{k}-\bm{q}_1-\bm{q}_2|) \delta P_0 (q_1) \nonumber \\
=& e^{6\eta} e^{-2 \alpha_k} \int \frac{q^2 \mathrm{d}q}{2 \pi^2} S^{(3)}_{ab} (q, k) \delta P_0 (q) .
\end{align}
The kernel functions are defined as
\begin{align}
X^{(2)}_{ab} (q, k) &= \frac{1}{2} \int_{-1}^{1} \mathrm{d} \mu_q \,
F^{(2)}_a (\bm{q}, \bm{k}-\bm{q}) F^{(2)}_b (\bm{q}, \bm{k}-\bm{q})
\Pfid_0 \left( \sqrt{k^2 - 2 k q \mu_q + q^2} \right) , \\
Y^{(2)}_{ab} (q, k) &= \frac{1}{2} \int_{-1}^{1} \mathrm{d} \mu_q \,
F^{(2)}_a (\bm{q}, \bm{k}-\bm{q})  \bar{\Gamma}^{(2)}_{b, \text{1-loop}} (\bm{q}, \bm{k}-\bm{q})
\Pfid_0 \left( \sqrt{k^2 - 2 k q \mu_q + q^2} \right) , \\
Z^{(2)}_{ab} (q, k) &= \frac{1}{2} \int_{-1}^{1} \mathrm{d} \mu_q \,
\bar{\Gamma}^{(2)}_{a, \text{1-loop}} (\bm{q}, \bm{k}-\bm{q}) \bar{\Gamma}^{(2)}_{b, \text{1-loop}} (\bm{q}, \bm{k}-\bm{q})
\Pfid_0 \left( \sqrt{k^2 - 2 k q \mu_q + q^2} \right) , \\
Q^{(2)}_{ab} (q, k) &= \int \frac{\mathrm{d}^3 \bm{p}}{(2\pi)^3} F^{(2)}_a (\bm{p}, \bm{k}-\bm{p}) K_b (q; p, |\bm{k}-\bm{p}|, k)
\Pfid_0 (|\bm{k}-\bm{p}|) \Pfid_0 (p) , \\
R^{(2)}_{ab} (q, k) &= \int \frac{\mathrm{d}^3 \bm{p}}{(2\pi)^3} \bar{\Gamma}^{(2)}_{a, \text{1-loop}} (\bm{p}, \bm{k}-\bm{p}) K_b (q; p, |\bm{k}-\bm{p}|, k)
\Pfid_0 (|\bm{k}-\bm{p}|) \Pfid_0 (p) , \\
S^{(3)}_{ab} (q, k) &= \frac{1}{2} \int_{-1}^{1} \mathrm{d}\mu_q \int \frac{\mathrm{d}^3 \bm{p}}{(2\pi)^3}
F^{(3)}_a (\bm{p}, \bm{q}, \bm{k}-\bm{p}-\bm{q}) F^{(3)}_b (\bm{p}, \bm{q}, \bm{k}-\bm{p}-\bm{q})
\Pfid_0 (|\bm{k}-\bm{p}-\bm{q}|) \Pfid_0 (p) .
\end{align}
In practical implementation, required kernels for power spectrum
are $L_a^{(1)}$, $M_a^{(1)}$, $X_{ab}^{(2)}$,
$Y_{ab}^{(2)}$, $Z_{ab}^{(2)}$, $Q_{ab}^{(2)}$, $R_{ab}^{(2)}$, $S_{ab}^{(3)}$
and they are tabulated with respect to arguments $(k, q)$.
As our default setting, the dimension and range of $k$ and $q$ is
$n_k = 120$ in $[10^{-3}, 1] \, \hMpcinv$
and $n_q = 200$ in $[5 \times 10^{-4}, 10] \, \hMpcinv$, respectively.
The integrals are calculated with Gaussian quadrature.
In addition to these kernels, we need to store diagram data for the fiducial cosmology
to compute the fiducial power spectra and the dimension and range of the wave-vector
is the same as the one of kernels.

\subsection{Kernels for bispectra}
The correction term for bispectra (Eq.~\ref{eq:rec_bk}) can also be decomposed into four terms:
\begin{equation}
\delta B_{abc} (\bm{k}_1, \bm{k}_2, \bm{k}_3) =
\delta B^\text{I,1}_{abc} (\bm{k}_1, \bm{k}_2, \bm{k}_3) +
\delta B^\text{I,2}_{abc} (\bm{k}_1, \bm{k}_2, \bm{k}_3) +
\delta B^\text{II}_{abc} (\bm{k}_1, \bm{k}_2, \bm{k}_3) +
\delta B^\text{III}_{abc} (\bm{k}_1, \bm{k}_2, \bm{k}_3) .
\end{equation}
Note that only for the $B^\text{I}_{abc}$ part, we include correction terms up to the next-to-leading order,
which corresponds to the $\delta B^\text{I,2}_{abc}$ term.
Without this term, the accuracy at large scales would degrade but
the computational cost is still kept up to one-dimensional integrations.
Each term is given by
\begin{align}
\delta B^\text{I,1}_{abc} (\bm{k}_1, \bm{k}_2, \bm{k}_3) =&
2 \left[
\delta \Gamma^{(2)}_{a,\reg} (\bm{k}_2, \bm{k}_3) \Gamma^{(1)}_{b,\reg} (k_2)
\Gamma^{(1)}_{c,\reg} (k_3) \Pfid_0 (k_2) \Pfid_0 (k_3)
\right. \nonumber \\
& + \Gamma^{(2)}_{a,\reg} (\bm{k}_2, \bm{k}_3) \delta \Gamma^{(1)}_{b,\reg} (k_2)
\Gamma^{(1)}_{c,\reg} (k_3) \Pfid_0 (k_2) \Pfid_0 (k_3)
\nonumber \\
& + \Gamma^{(2)}_{a,\reg} (\bm{k}_2, \bm{k}_3) \Gamma^{(1)}_{b,\reg} (k_2)
\delta \Gamma^{(1)}_{c,\reg} (k_3) \Pfid_0 (k_2) \Pfid_0 (k_3)
\nonumber \\
& + \Gamma^{(2)}_{a,\reg} (\bm{k}_2, \bm{k}_3) \Gamma^{(1)}_{b,\reg} (k_2)
\Gamma^{(1)}_{c,\reg} (k_3) \delta P_0 (k_2) \Pfid_0 (k_3)
\nonumber \\
& \left. + \Gamma^{(2)}_{a,\reg} (\bm{k}_2, \bm{k}_3) \Gamma^{(1)}_{b,\reg} (k_2)
\Gamma^{(1)}_{c,\reg} (k_3) \Pfid_0 (k_2) \delta P_0 (k_3) \right]
\nonumber \\
& + [\text{perm.}: (a, b, c) \to (b, c, a), (\bm{k}_1, \bm{k}_2, \bm{k}_3) \to (\bm{k}_2, \bm{k}_3, \bm{k}_1)]
\nonumber \\
& + [\text{perm.}: (a, b, c) \to (c, a, b), (\bm{k}_1, \bm{k}_2, \bm{k}_3) \to (\bm{k}_3, \bm{k}_1, \bm{k}_2)] ,
\end{align}
\begin{align}
\delta B^\text{I,2}_{abc} (\bm{k}_1, \bm{k}_2, \bm{k}_3) =&
2 \left[ \delta \Gamma^{(2)}_{a,\reg} (\bm{k}_2, \bm{k}_3) \delta \Gamma^{(1)}_{b,\reg} (k_2)
\Gamma^{(1)}_{c,\reg} (k_3) \Pfid_0 (k_2) \Pfid_0 (k_3)
\right. \nonumber \\
& + \delta \Gamma^{(2)}_{a,\reg} (\bm{k}_2, \bm{k}_3) \Gamma^{(1)}_{b,\reg} (k_2)
\delta \Gamma^{(1)}_{c,\reg} (k_3) \Pfid_0 (k_2) \Pfid_0 (k_3)
\nonumber \\
& + \delta \Gamma^{(2)}_{a,\reg} (\bm{k}_2, \bm{k}_3) \Gamma^{(1)}_{b,\reg} (k_2)
\Gamma^{(1)}_{c,\reg} (k_3) \delta P_0 (k_2) \Pfid_0 (k_3)
\nonumber \\
& + \delta \Gamma^{(2)}_{a,\reg} (\bm{k}_2, \bm{k}_3) \Gamma^{(1)}_{b,\reg} (k_2)
\Gamma^{(1)}_{c,\reg} (k_3) \Pfid_0 (k_2) \delta P_0 (k_3)
\nonumber \\
& + \Gamma^{(2)}_{a,\reg} (\bm{k}_2, \bm{k}_3) \delta \Gamma^{(1)}_{b,\reg} (k_2)
\delta \Gamma^{(1)}_{c,\reg} (k_3) \Pfid_0 (k_2) \Pfid_0 (k_3)
\nonumber \\
& + \Gamma^{(2)}_{a,\reg} (\bm{k}_2, \bm{k}_3) \delta \Gamma^{(1)}_{b,\reg} (k_2)
\Gamma^{(1)}_{c,\reg} (k_3) \delta P_0 (k_2) \Pfid_0 (k_3)
\nonumber \\
& + \Gamma^{(2)}_{a,\reg} (\bm{k}_2, \bm{k}_3) \delta \Gamma^{(1)}_{b,\reg} (k_2)
\Gamma^{(1)}_{c,\reg} (k_3) \Pfid_0 (k_2) \delta P_0 (k_3)
\nonumber \\
& + \Gamma^{(2)}_{a,\reg} (\bm{k}_2, \bm{k}_3) \Gamma^{(1)}_{b,\reg} (k_2)
\delta \Gamma^{(1)}_{c,\reg} (k_3) \Pfid_0 (k_2) \Pfid_0 (k_3)
\nonumber \\
& + \Gamma^{(2)}_{a,\reg} (\bm{k}_2, \bm{k}_3) \Gamma^{(1)}_{b,\reg} (k_2)
\delta \Gamma^{(1)}_{c,\reg} (k_3) \delta P_0 (k_2) \Pfid_0 (k_3)
\nonumber \\
& + \Gamma^{(2)}_{a,\reg} (\bm{k}_2, \bm{k}_3) \Gamma^{(1)}_{b,\reg} (k_2)
\delta \Gamma^{(1)}_{c,\reg} (k_3) \Pfid_0 (k_2) \delta P_0 (k_3)
\nonumber \\
& \left. + \Gamma^{(2)}_{a,\reg} (\bm{k}_2, \bm{k}_3) \Gamma^{(1)}_{b,\reg} (k_2)
\Gamma^{(1)}_{c,\reg} (k_3) \delta P_0 (k_2) \delta P_0 (k_3) \right]
\nonumber \\
& + [\text{perm.}: (a, b, c) \to (b, c, a), (\bm{k}_1, \bm{k}_2, \bm{k}_3) \to (\bm{k}_2, \bm{k}_3, \bm{k}_1)]
\nonumber \\
& + [\text{perm.}: (a, b, c) \to (c, a, b), (\bm{k}_1, \bm{k}_2, \bm{k}_3) \to (\bm{k}_3, \bm{k}_1, \bm{k}_2)] ,
\end{align}
\begin{align}
\delta B^\text{II}_{abc} (\bm{k}_1, \bm{k}_2, \bm{k}_3) =&
8 \int \frac{\mathrm{d}^3 \bm{q}}{(2\pi)^3}  \delta P_0 (q) \times \nonumber \\
& \left[
\Gamma^{(2)}_{a,\reg} (\bm{k}_1-\bm{q}, \bm{q})
\Gamma^{(2)}_{b,\reg} (\bm{k}_2+\bm{q}, -\bm{q})
\Gamma^{(2)}_{c,\reg} (-\bm{q}-\bm{k}_2, -\bm{k}_1+\bm{q})
\Pfid_0 (|\bm{k}_1 - \bm{q}|) \Pfid_0 (|\bm{k}_2 + \bm{q}|)
\right. \nonumber \\
& + \Gamma^{(2)}_{a,\reg} (\bm{k}_1-\bm{q}, \bm{q})
\Gamma^{(2)}_{b,\reg} (-\bm{q}-\bm{k}_3, -\bm{k}_1+\bm{q})
\Gamma^{(2)}_{c,\reg} (\bm{k}_3+\bm{q}, -\bm{q})
\Pfid_0 (|\bm{k}_1 - \bm{q}|) \Pfid_0 (|\bm{k}_3 + \bm{q}|)
\nonumber \\
& \left. + \Gamma^{(2)}_{a,\reg} (-\bm{q}-\bm{k}_3, -\bm{k}_2+\bm{q})
\Gamma^{(2)}_{b,\reg} (\bm{q}, \bm{k}_2-\bm{q})
\Gamma^{(2)}_{c,\reg} (\bm{k}_3+\bm{q}, -\bm{q})
\Pfid_0 (|\bm{k}_2 - \bm{q}|) \Pfid_0 (|\bm{k}_3 + \bm{q}|) \right] ,
\end{align}
\begin{align}
\delta B^\text{III}_{abc} (\bm{k}_1, \bm{k}_2, \bm{k}_3) =&
6 \int \frac{\mathrm{d}^3 \bm{q}}{(2\pi)^3} \times \nonumber \\
& \left[ \Gamma^{(3)}_{a,\reg} (-\bm{k}_3, -\bm{k}_2+\bm{q}, -\bm{q})
\Gamma^{(2)}_{b,\reg} (\bm{k}_2-\bm{q}, \bm{q})
\Gamma^{(1)}_{c,\reg} (k_3) \right. \nonumber \\
& \times \Pfid_0 (|\bm{k}_2-\bm{q}|)
( 2 \Pfid_0 (k_3) \delta P_0 (q) + \delta P_0 (k_3) \Pfid_0 (q) )
\nonumber \\
& + \Gamma^{(3)}_{a,\reg} (-\bm{k}_2, -\bm{k}_3+\bm{q}, -\bm{q})
\Gamma^{(2)}_{c,\reg} (\bm{k}_3-\bm{q}, \bm{q})
\Gamma^{(1)}_{b,\reg} (k_2) \nonumber \\
& \times \Pfid_0 (|\bm{k}_3-\bm{q}|)
( 2 \Pfid_0 (k_2) \delta P_0 (q) + \delta P_0 (k_2) \Pfid_0 (q) )
\nonumber \\
& + \Gamma^{(3)}_{b,\reg} (-\bm{q}, -\bm{k}_1+\bm{q}, -\bm{k}_3)
\Gamma^{(2)}_{a,\reg} (\bm{k}_1-\bm{q}, \bm{q})
\Gamma^{(1)}_{c,\reg} (k_3) \nonumber \\
& \times \Pfid_0 (|\bm{k}_1-\bm{q}|)
( 2 \Pfid_0 (k_3) \delta P_0 (q) + \delta P_0 (k_3) \Pfid_0 (q) )
\nonumber \\
& + \Gamma^{(3)}_{b,\reg} (-\bm{k}_1, -\bm{k}_3+\bm{q}, -\bm{q})
\Gamma^{(2)}_{c,\reg} (\bm{k}_3-\bm{q}, \bm{q})
\Gamma^{(1)}_{a,\reg} (k_1) \nonumber \\
& \times \Pfid_0 (|\bm{k}_3-\bm{q}|)
( 2 \Pfid_0 (k_1) \delta P_0 (q) + \delta P_0 (k_1) \Pfid_0 (q) )
\nonumber \\
& + \Gamma^{(3)}_{c,\reg} (-\bm{k}_1+\bm{q}, -\bm{q}, -\bm{k}_2)
\Gamma^{(2)}_{a,\reg} (\bm{k}_1-\bm{q}, \bm{q})
\Gamma^{(1)}_{b,\reg} (k_2) \nonumber \\
& \times \Pfid_0 (|\bm{k}_1-\bm{q}|)
( 2 \Pfid_0 (k_2) \delta P_0 (q) + \delta P_0 (k_2) \Pfid_0 (q) )
\nonumber \\
& + \Gamma^{(3)}_{c,\reg} (-\bm{k}_1, -\bm{q}, -\bm{k}_2+\bm{q})
\Gamma^{(2)}_{b,\reg} (\bm{k}_2-\bm{q}, \bm{q})
\Gamma^{(1)}_{a,\reg} (k_1) \nonumber \\
& \left. \times \Pfid_0 (|\bm{k}_2-\bm{q}|)
( 2 \Pfid_0 (k_1) \delta P_0 (q) + \delta P_0 (k_1) \Pfid_0 (q) ) \right] .
\end{align}

For $\delta \Gamma^{(1)}_{a,\reg} (k)$ and $\delta \Gamma^{(1)}_{a,\reg} (\bm{k}_1, \bm{k}_2)$
in $\delta B^\text{I}_{abc} (\bm{k}_1, \bm{k}_2, \bm{k}_3)$,
the correction at 1-loop order is given as
\begin{align}
\delta \Gamma^{(1)}_{a,\reg} (k; \eta) &= e^{3 \eta} e^{-\alpha_k}
\delta \bar{\Gamma}^{(1)}_{a, \text{1-loop}} (k) , \\
\delta \Gamma^{(2)}_{a,\reg} (\bm{k}_1, \bm{k}_2; \eta) &=
e^{4 \eta} e^{-\alpha_k} \delta \bar{\Gamma}^{(2)}_{a, \text{1-loop}} (\bm{k}_1, \bm{k}_2) .
\end{align}

We rewrite $\delta B^\text{II}_{abc}$ and $\delta B^\text{III}_{abc}$ in the similar form as in power spectrum
and define kernel functions as follows.
\begin{align}
\delta B^{\text{II}}_{abc}  (\bm{k}_1, \bm{k}_2, \bm{k}_3) =&
8 e^{6\eta} e^{- \alpha_{k_{123}}}
\int \frac{q^2 \mathrm{d}q}{2 \pi^2} T^{(3)}_{abc} (q; \bm{k}_1, \bm{k}_2, \bm{k}_3) \delta P_0 (q) , \\
\delta B^{\text{III}}_{abc}  (\bm{k}_1, \bm{k}_2, \bm{k}_3) =&
12 e^{6\eta} e^{- \alpha_{k_{123}}}
\int \frac{q^2 \mathrm{d}q}{2 \pi^2} U^{(3)}_{abc} (q; \bm{k}_1, \bm{k}_2, \bm{k}_3) \delta P_0 (q)
\nonumber \\
& + 6 e^{6\eta} e^{- \alpha_{k_{123}}}
\sum_{i=1}^3 V^{(3)}_{abc,i} (\bm{k}_1, \bm{k}_2, \bm{k}_3)  \delta P_0 (k_i) ,
\end{align}
where $\alpha_{k_{123}} \equiv \alpha_{k_1} + \alpha_{k_2} + \alpha_{k_3}$.

The kernel functions are
\begin{align}
T^{(3)}_{abc} (q; \bm{k}_1, \bm{k}_2, \bm{k}_3) =& \int \frac{\mathrm{d}^2 \bm{\Omega}_q}{4\pi} \times
\nonumber \\
& \left[ F^{(2)} _a (\bm{k}_1-\bm{q}, \bm{q}) F^{(2)}_b (\bm{k}_2+\bm{q}, -\bm{q})
F^{(2)}_c (-\bm{q}-\bm{k}_2, -\bm{k}_1+\bm{q}) \Pfid_0 (|\bm{k}_1 - \bm{q}|) \Pfid_0 (|\bm{k}_2 + \bm{q}|)
\right. \nonumber \\
& + F^{(2)} _a (\bm{k}_1-\bm{q}, \bm{q}) F^{(2)}_b (-\bm{q}-\bm{k}_3, -\bm{k}_1+\bm{q})
F^{(2)}_c (\bm{k}_3+\bm{q}, -\bm{q}) \Pfid_0 (|\bm{k}_1 - \bm{q}|) \Pfid_0 (|\bm{k}_3 + \bm{q}|)
\nonumber \\
& \left. + F^{(2)} _a (-\bm{q}-\bm{k}_3, -\bm{k}_2+\bm{q}) F^{(2)}_b (\bm{q}, \bm{k}_2-\bm{q})
F^{(2)}_c (\bm{k}_3+\bm{q}, -\bm{q}) \Pfid_0 (|\bm{k}_2 - \bm{q}|) \Pfid_0 (|\bm{k}_3 + \bm{q}|) \right] , \\
U^{(3)}_{abc} (q; \bm{k}_1, \bm{k}_2, \bm{k}_3) =& \int \frac{\mathrm{d}^2 \bm{\Omega}_q}{4\pi} \times \nonumber \\
& \left[ F^{(3)}_a (-\bm{k}_3, -\bm{k}_2+\bm{q}, -\bm{q}) F^{(2)}_b (\bm{k}_2-\bm{q}, \bm{q})
\Pfid_0 (|\bm{k}_2-\bm{q}|) \Pfid_0 (k_3) \right. \nonumber \\
& + F^{(3)}_a (-\bm{k}_2, -\bm{k}_3+\bm{q}, -\bm{q}) F^{(2)}_c (\bm{k}_3-\bm{q}, \bm{q})
\Pfid_0 (|\bm{k}_3-\bm{q}|) \Pfid_0 (k_2) \nonumber \\
& + F^{(3)}_b (-\bm{q}, -\bm{k}_1+\bm{q}, -\bm{k}_3) F^{(2)}_a (\bm{k}_1-\bm{q}, \bm{q})
\Pfid_0 (|\bm{k}_1-\bm{q}|) \Pfid_0 (k_3) \nonumber \\
& + F^{(3)}_b (-\bm{k}_1, -\bm{k}_3+\bm{q}, -\bm{q}) F^{(2)}_c (\bm{k}_3-\bm{q}, \bm{q})
\Pfid_0 (|\bm{k}_3-\bm{q}|) \Pfid_0 (k_1) \nonumber \\
& + F^{(3)}_c (-\bm{k}_1+\bm{q}, -\bm{q}, -\bm{k}_2) F^{(2)}_a (\bm{k}_1-\bm{q}, \bm{q})
\Pfid_0 (|\bm{k}_1-\bm{q}|) \Pfid_0 (k_2) \nonumber \\
& \left. + F^{(3)}_c (-\bm{k}_1, -\bm{q}, -\bm{k}_2+\bm{q}) F^{(2)}_b (\bm{k}_2-\bm{q}, \bm{q})
\Pfid_0 (|\bm{k}_2-\bm{q}|) \Pfid_0 (k_1) \right] , \\
V^{(3)}_{abc,1} (\bm{k}_1, \bm{k}_2, \bm{k}_3) =& \int \frac{\mathrm{d}^3 \bm{q}}{(2 \pi)^3} \times \nonumber \\
& \left[ F^{(3)}_b (-\bm{k}_1, -\bm{k}_3+\bm{q}, -\bm{q}) F^{(2)}_c (\bm{k}_3-\bm{q}, \bm{q})
\Pfid_0 (|\bm{k}_3-\bm{q}|) \Pfid_0 (q) \right. \nonumber \\
& \left. + F^{(3)}_c (-\bm{k}_1, -\bm{q}, -\bm{k}_2+\bm{q}) F^{(2)}_b (\bm{k}_2-\bm{q}, \bm{q})
\Pfid_0 (|\bm{k}_2-\bm{q}|) \Pfid_0 (q) \right] , \\
V^{(3)}_{abc,2} (\bm{k}_1, \bm{k}_2, \bm{k}_3) =& \int \frac{\mathrm{d}^3 \bm{q}}{(2 \pi)^3} \times \nonumber \\
& \left[ F^{(3)}_a (-\bm{k}_2, -\bm{k}_3+\bm{q}, -\bm{q}) F^{(2)}_c (\bm{k}_3-\bm{q}, \bm{q})
\Pfid_0 (|\bm{k}_3-\bm{q}|) \Pfid_0 (q) \right. \nonumber \\
& \left. + F^{(3)}_c (-\bm{k}_1+\bm{q}, -\bm{q}, -\bm{k}_2) F^{(2)}_a (\bm{k}_1-\bm{q}, \bm{q})
\Pfid_0 (|\bm{k}_1-\bm{q}|) \Pfid_0 (q) \right] , \\
V^{(3)}_{abc,3} (\bm{k}_1, \bm{k}_2, \bm{k}_3) =& \int \frac{\mathrm{d}^3 \bm{q}}{(2 \pi)^3} \times \nonumber \\
& \left[ F^{(3)}_a (-\bm{k}_3, -\bm{k}_2+\bm{q}, -\bm{q}) F^{(2)}_b (\bm{k}_2-\bm{q}, \bm{q})
\Pfid_0 (|\bm{k}_2-\bm{q}|) \Pfid_0 (q) \right. \nonumber \\
& \left. + F^{(3)}_b (-\bm{q}, -\bm{k}_1+\bm{q}, -\bm{k}_3) F^{(2)}_a (\bm{k}_1-\bm{q}, \bm{q})
\Pfid_0 (|\bm{k}_1-\bm{q}|) \Pfid_0 (q) \right] .
\end{align}
For bispectrum, $T^{(3)}_{abc}$, $U^{(3)}_{abc}$, $V^{(3)}_{abc,i}$
are stored to compute correction terms
and the dimension and range of $q$ is the same as the case in power spectrum.
We sample triangles $(k_1, k_2, k_3)$ by log-equally sampling with respect to $(K_1, K_2, K_3)$,
which is defined in Eq.~\eqref{eq:def_K},
in the range $[10^{-3}, 0.6] \, h \, \mathrm{Mpc}^{-1}$
with the number of sampling $n_K = 100$ in each dimension.

\subsection{The scaling relation for kernels}
As described in Section~\ref{sec:distance}, power spectrum is rescaled so that
the distance between target and fiducial models is minimized.
The propagators and kernels involve linear power spectrum,
and thus, we need to scale them as well.
The scaling for propagators is given as
\begin{equation}
  \bar{\Gamma}^{(1)}_{a, \text{1-loop}} (k) \to c \bar{\Gamma}^{(1)}_{a, \text{1-loop}} (k) , \
  \bar{\Gamma}^{(1)}_{a, \text{2-loop}} (k) \to c^2 \bar{\Gamma}^{(1)}_{a, \text{2-loop}} (k) , \
  \bar{\Gamma}^{(2)}_{a, \text{1-loop}} (\bm{k}_1, \bm{k}_2) \to
  c \bar{\Gamma}^{(2)}_{a, \text{1-loop}} (\bm{k}_1, \bm{k}_2) ,
\end{equation}
and scaling for kernels is given as
\begin{gather}
  L^{(1)}_a (q; k) \to L^{(1)}_a (q; k) , \
  M^{(1)}_a (q; k) \to c M^{(1)}_a (q; k), \
  N^{(2)}_a (q; k_1, k_2, k_3) \to N^{(2)}_a (q; k_1, k_2, k_3), \\
  X^{(2)}_{ab} (q, k) \to c X^{(2)}_{ab} (q, k) , \
  Y^{(2)}_{ab} (q, k) \to c^2 Y^{(2)}_{ab} (q, k) , \
  Z^{(2)}_{ab} (q, k) \to c^3 Z^{(2)}_{ab} (q, k) , \\
  Q^{(2)}_{ab} (q, k) \to c^2 Q^{(2)}_{ab} (q, k) , \
  R^{(2)}_{ab} (q, k) \to c^3 R^{(2)}_{ab} (q, k) , \
  S^{(3)}_{ab} (q, k) \to c^2 S^{(3)}_{ab} (q, k) , \\
  T^{(3)}_{abc} (q; \bm{k}_1, \bm{k}_2, \bm{k}_3) \to
  c^2 T^{(3)}_{abc} (q; \bm{k}_1, \bm{k}_2, \bm{k}_3) , \
  U^{(3)}_{abc} (q; \bm{k}_1, \bm{k}_2, \bm{k}_3) \to
  c^2 U^{(3)}_{abc} (q; \bm{k}_1, \bm{k}_2, \bm{k}_3) , \nonumber \\
  V^{(3)}_{abc,i} (\bm{k}_1, \bm{k}_2, \bm{k}_3) \to
  c^2 V^{(3)}_{abc,i} (\bm{k}_1, \bm{k}_2, \bm{k}_3) \ (i=1,2,3).
\end{gather}

\section{The effect of UV cutoff scale}
\label{sec:UV_cutoff}
\begin{figure*}
  \includegraphics[width=0.75\columnwidth]{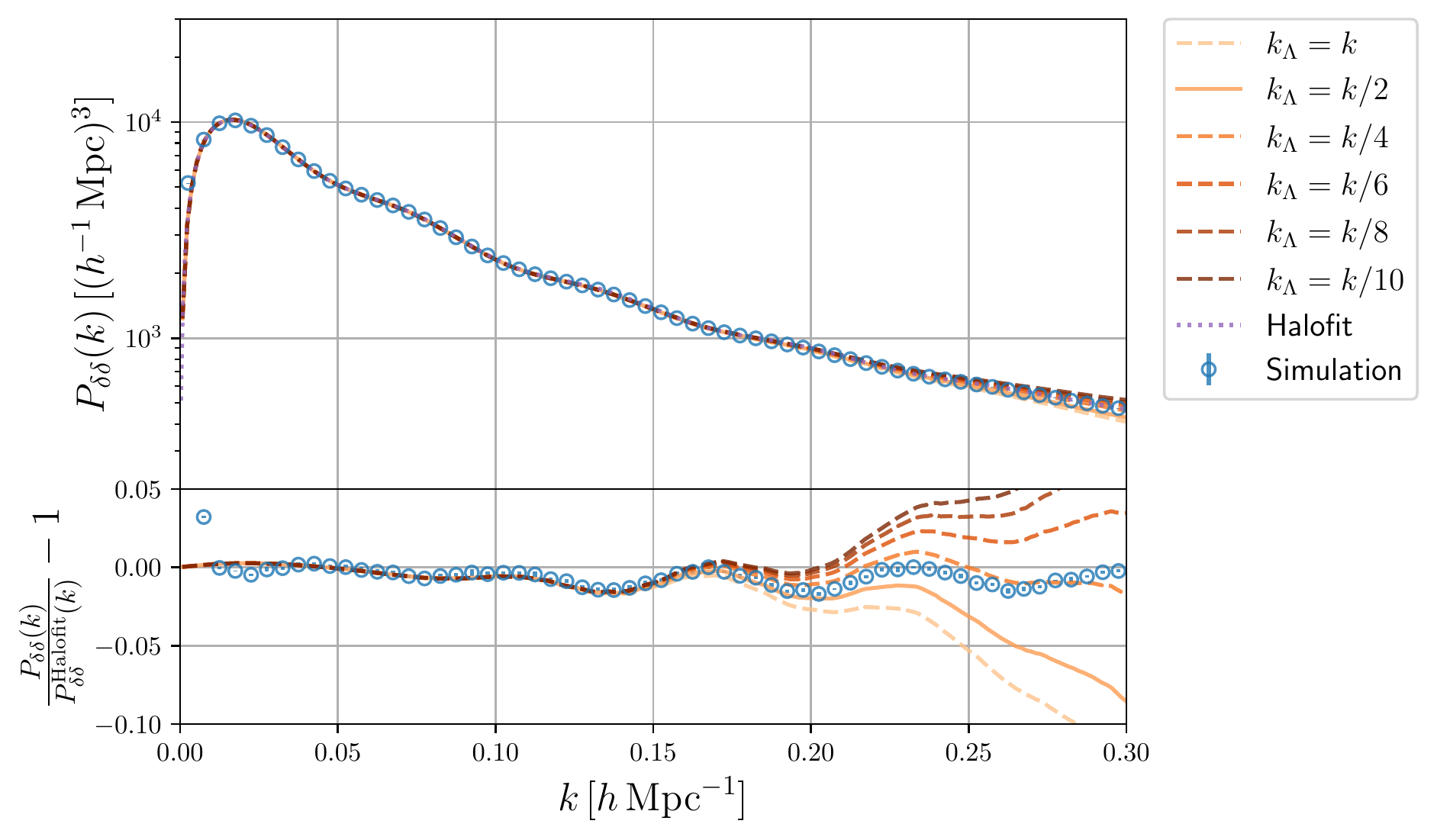}
  \caption{Power spectra based on RegPT
  with UV cutoff scales of $k_\Lambda = k, k/2, k/4, k/6, k/8, k/10$
  at the redshift $z = 0.901$.
  The results of $N$-body simulation and \textit{Halofit} are also shown.
  The fiducial result ($k_\Lambda = k/2$) is shown in solid line
  and other results are shown in dashed lines.}
  \label{fig:Pk_lambda}
\end{figure*}

\begin{figure*}
  \includegraphics[width=0.75\columnwidth]{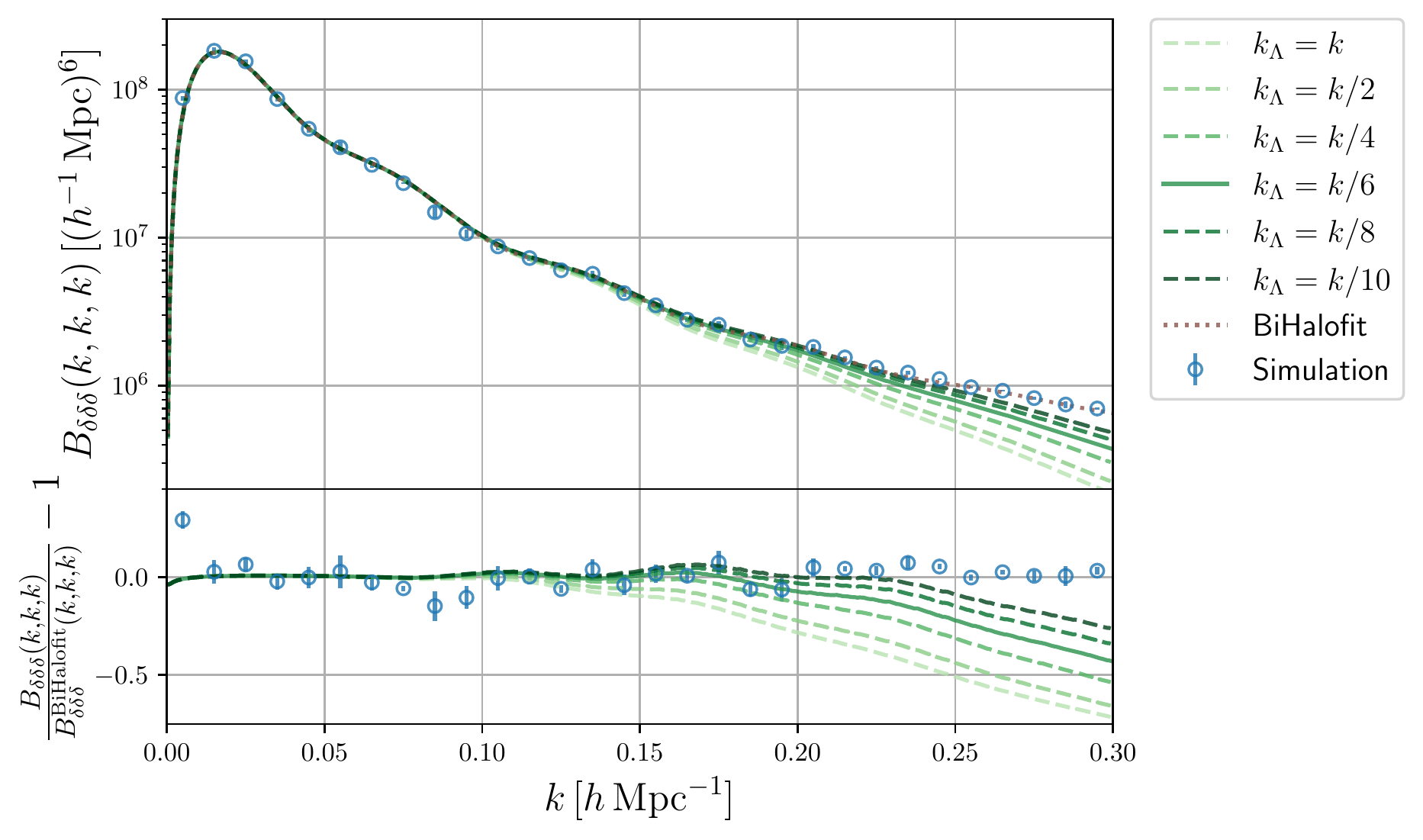}
  \caption{Bispectra for equilateral configurations
  based on RegPT with UV cutoff scales of $k_\Lambda = k, k/2, k/4, k/6, k/8, k/10$
  at the redshift $z = 0.901$.
  The results of $N$-body simulation and \textit{BiHalofit} are also shown.
  The fiducial result ($k_\Lambda = k/6$) is shown in solid line
  and other results are shown in dashed lines.}
  \label{fig:Bk_equil_lambda}
\end{figure*}

\begin{figure*}
  \includegraphics[width=0.75\columnwidth]{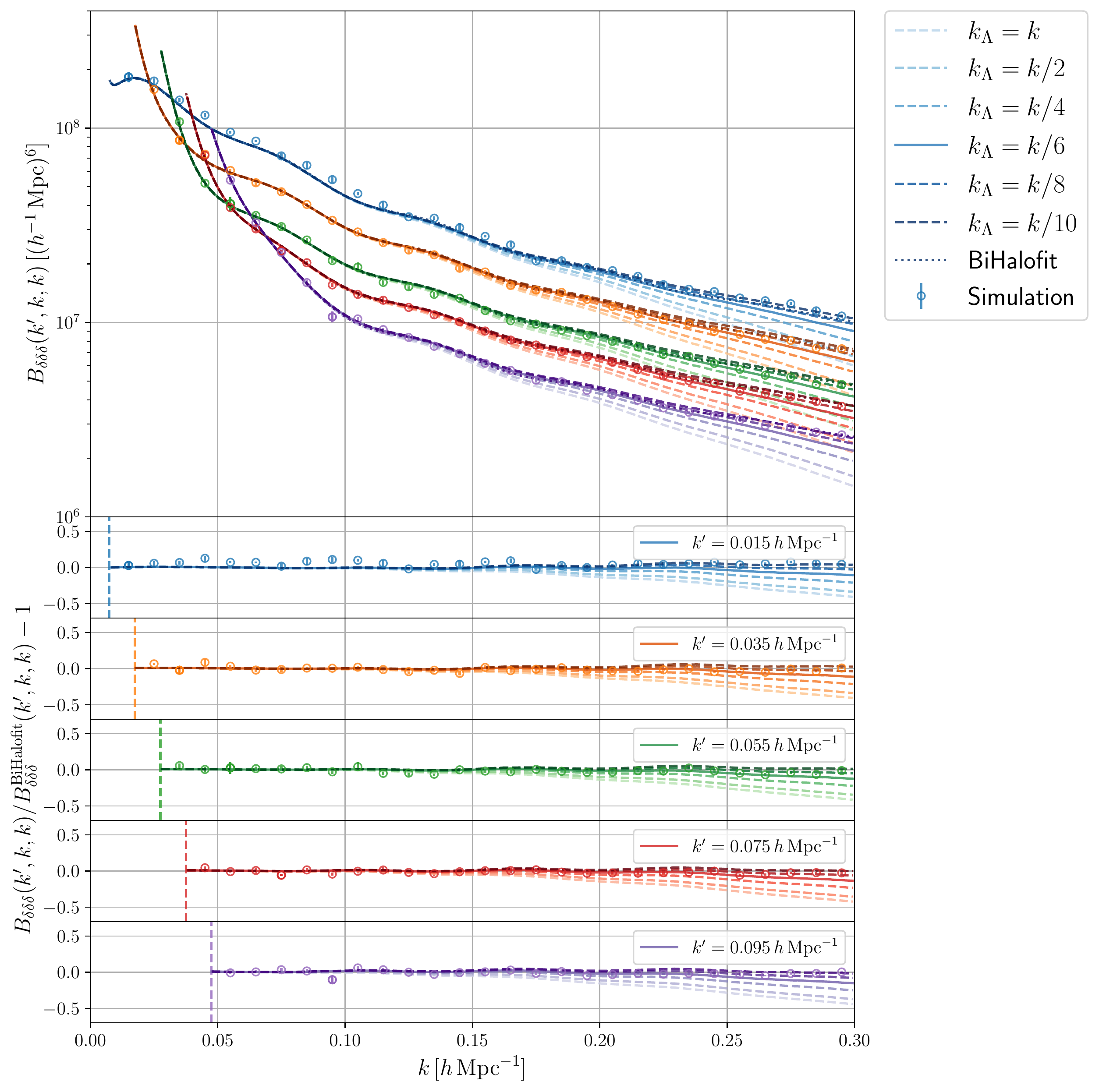}
  \caption{Bispectra for isosceles configurations $(k', k, k)$
  based on RegPT with UV cutoff scales of $k_\Lambda = k, k/2, k/4, k/6, k/8, k/10$
  at the redshift $z = 0.901$.
  The results of $N$-body simulation and \textit{BiHalofit} are also shown.
  The fiducial result ($k_\Lambda = k/6$) is shown in solid line
  and other results are shown in dashed lines.}
  \label{fig:Bk_iso1_lambda}
\end{figure*}

\begin{figure*}
  \includegraphics[width=0.75\columnwidth]{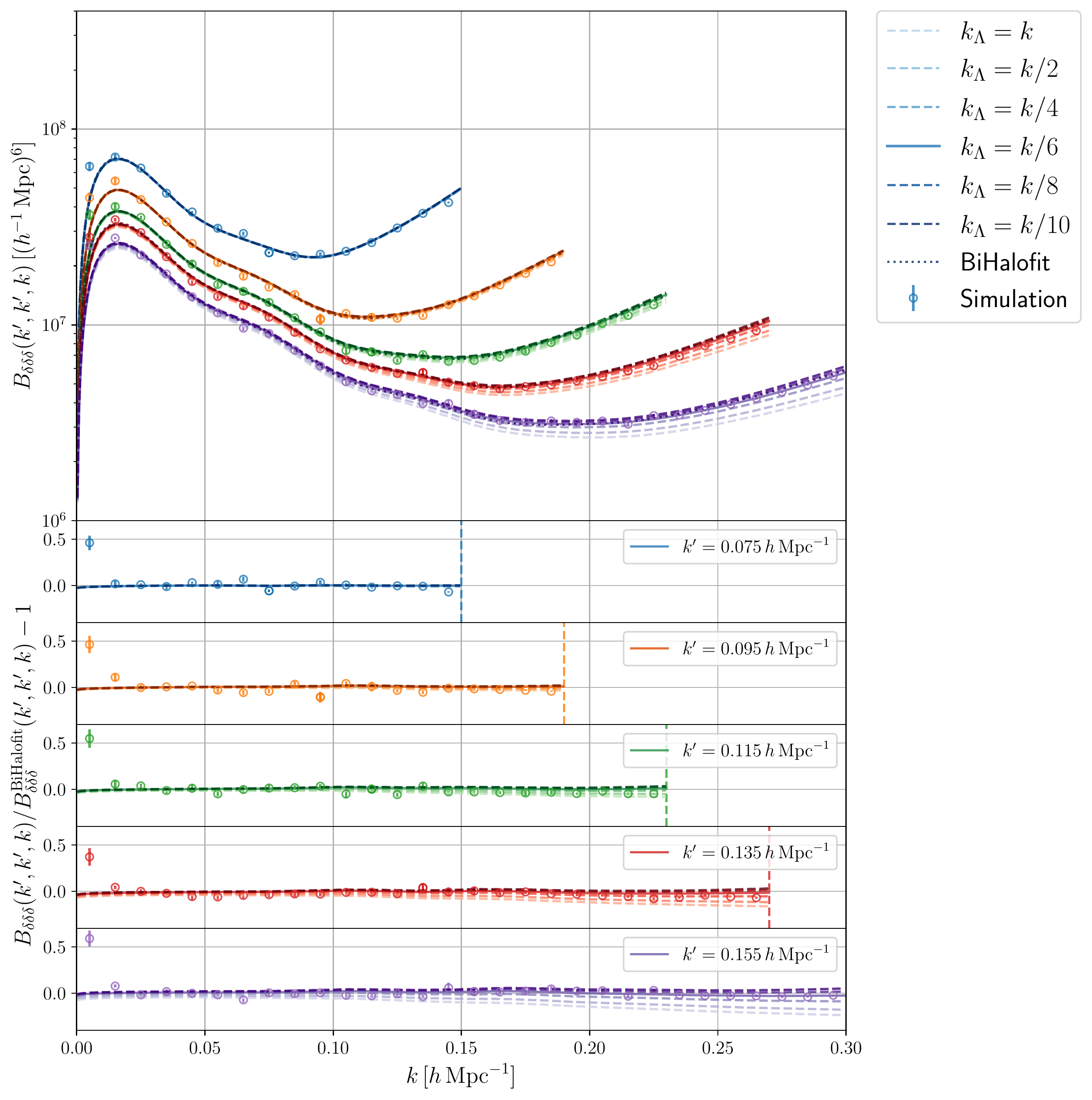}
  \caption{Bispectra for isosceles configurations $(k', k', k)$
  based on RegPT with UV cutoff scales of $k_\Lambda = k, k/2, k/4, k/6, k/8, k/10$
  at the redshift $z = 0.901$.
  The results of $N$-body simulation and \textit{BiHalofit} are also shown.
  The fiducial result ($k_\Lambda = k/6$) is shown in solid line
  and other results are shown in dashed lines.}
  \label{fig:Bk_iso2_lambda}
\end{figure*}

Here, we investigate how the UV cutoff scale of displacement dispersion (Eq.~\ref{eq:sigma_d})
affects power spectra and bispectra based on RegPT.
In Figures~\ref{fig:Pk_lambda}, \ref{fig:Bk_equil_lambda}, \ref{fig:Bk_iso1_lambda},
and \ref{fig:Bk_iso2_lambda},
power spectra and bispectra with UV cutoff scales of $k_\Lambda = k, k/2, k/4, k/6, k/8, k/10$
at the redshift $z = 0.901$ are shown.
By comparing with the $N$-body simulation results,
we have adopted $k_\Lambda = k/2$ for power spectra
and $k_\Lambda = k/6$ for bispectra.

\section{Results for other redshifts}
\label{sec:other_z}
\begin{figure*}
  \includegraphics[width=\textwidth]{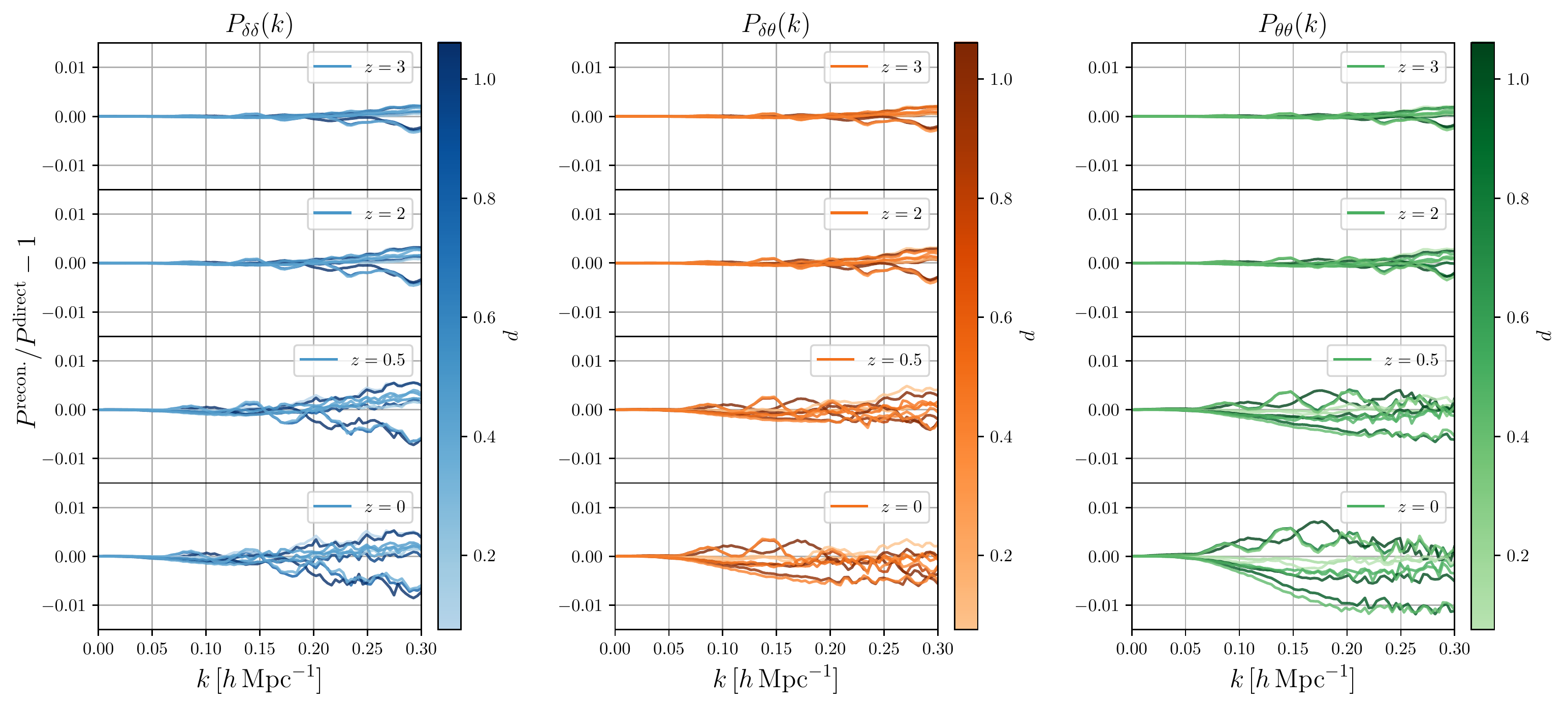}
  \caption{Same as Figure~\ref{fig:Pk_validation} but for $z = 3, 2, 0.5, 0$.}
  \label{fig:Pk_validation_z}
\end{figure*}

\begin{figure*}
  \includegraphics[width=\textwidth]{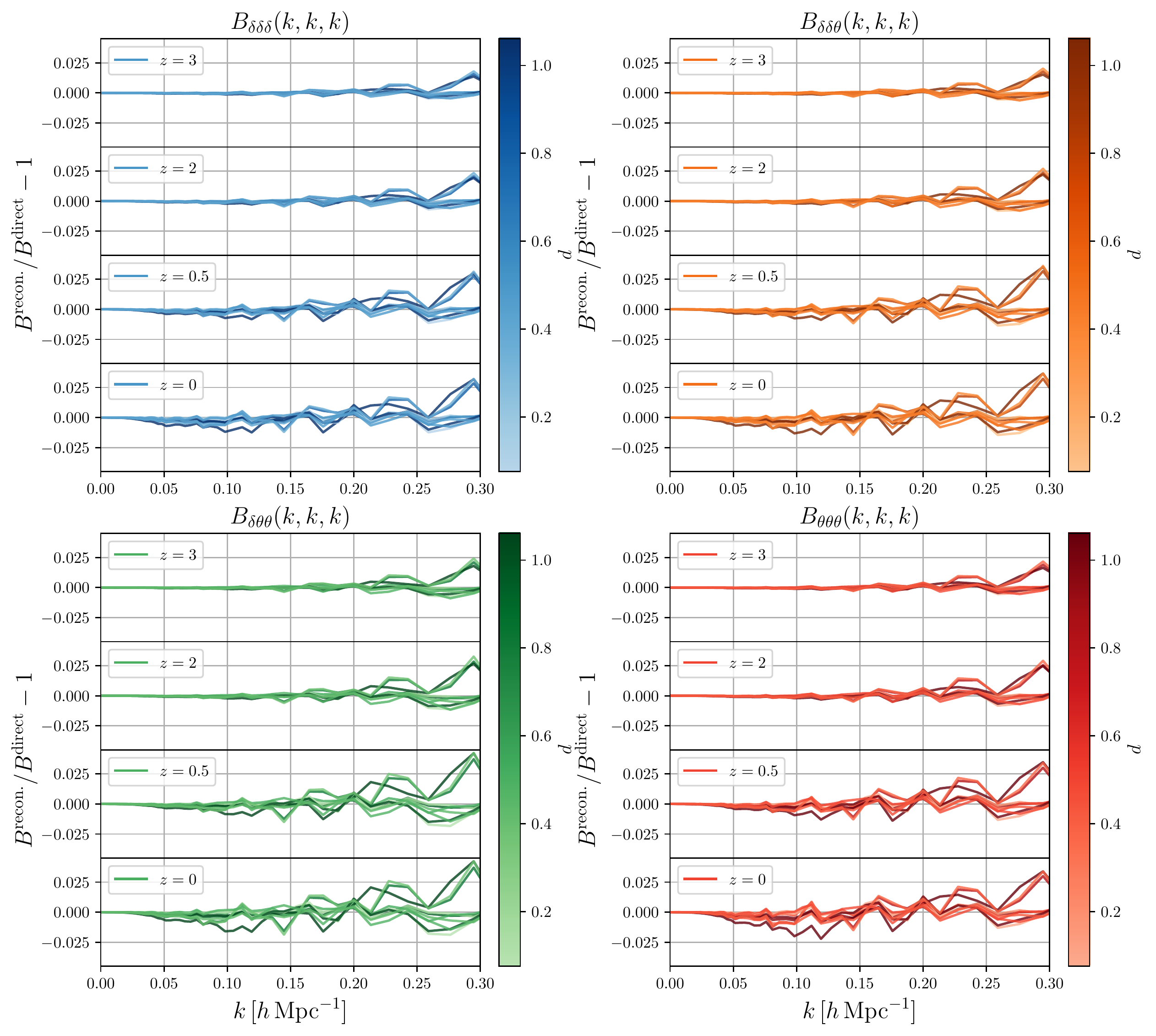}
  \caption{Same as Figure~\ref{fig:Bk_equil_validation} but for $z = 3, 2, 0.5, 0$.}
  \label{fig:Bk_equil_validation_z}
\end{figure*}

\begin{figure*}
  \includegraphics[width=\textwidth]{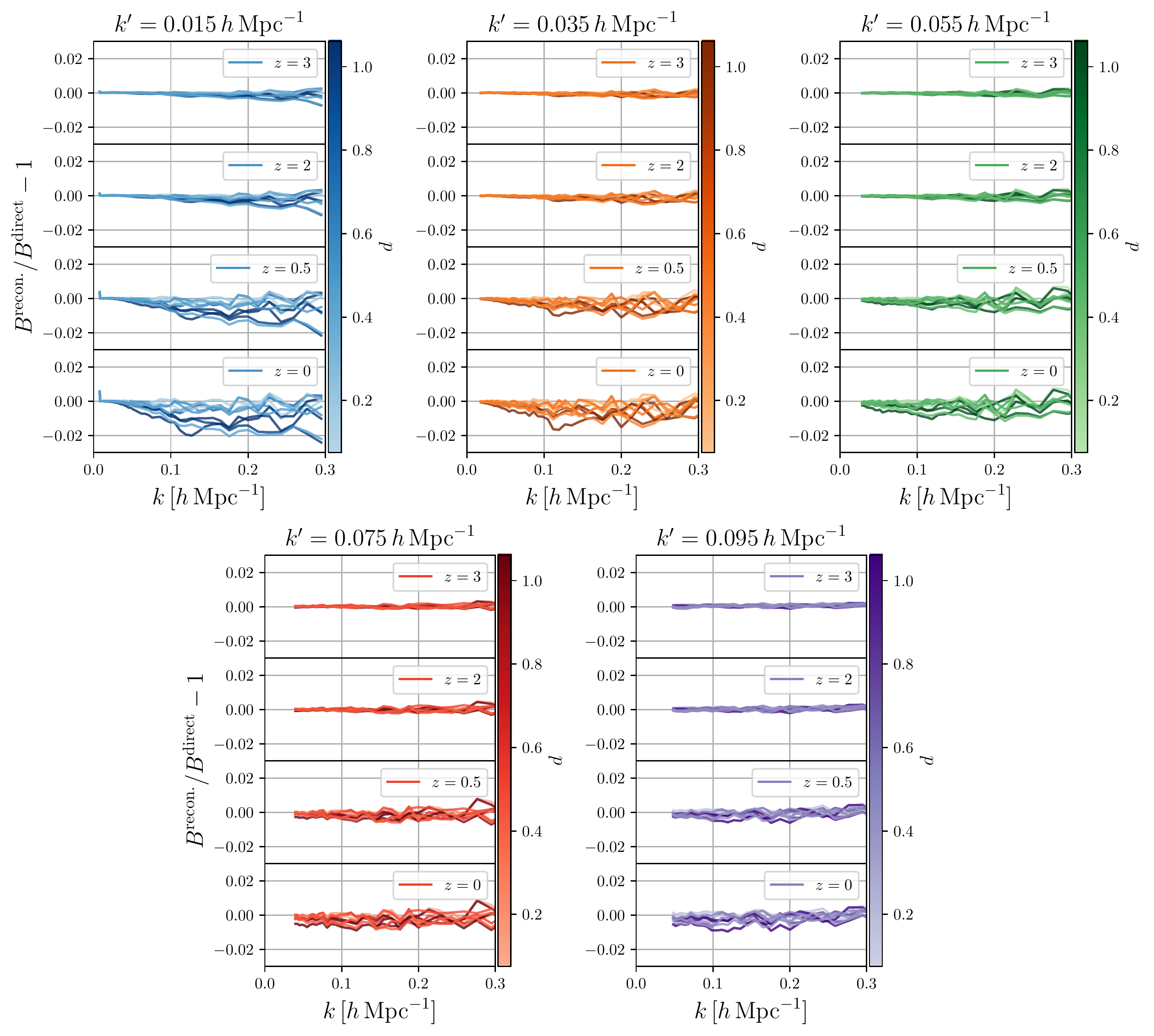}
  \caption{Same as Figure~\ref{fig:Bk_iso1_validation} but for $z = 3, 2, 0.5, 0$.}
  \label{fig:Bk_iso1_validation_z}
\end{figure*}

\begin{figure*}
  \includegraphics[width=\textwidth]{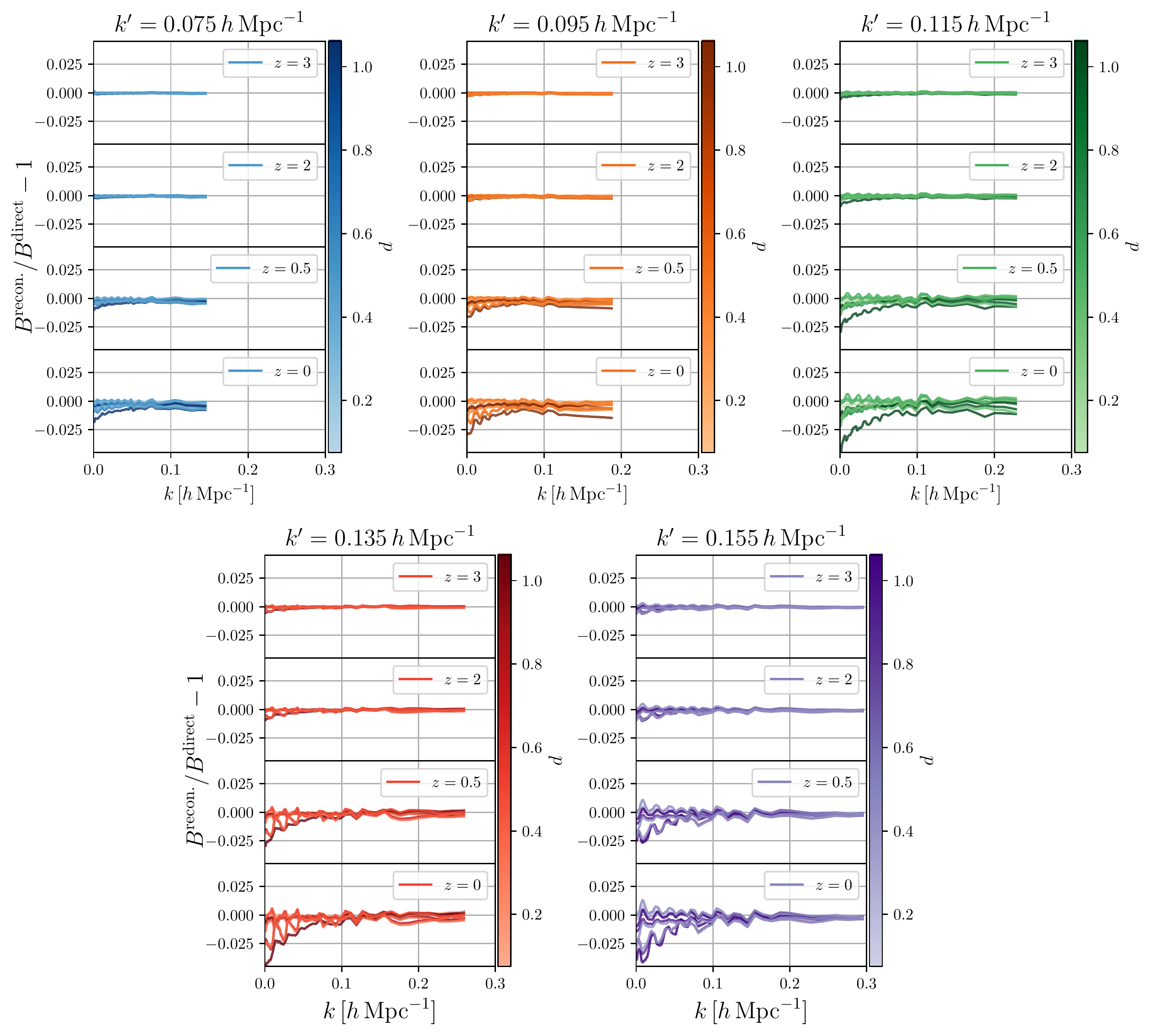}
  \caption{Same as Figure~\ref{fig:Bk_iso2_validation} but for $z = 3, 2, 0.5, 0$.}
  \label{fig:Bk_iso2_validation_z}
\end{figure*}

\begin{figure*}
  \includegraphics[width=0.75\columnwidth]{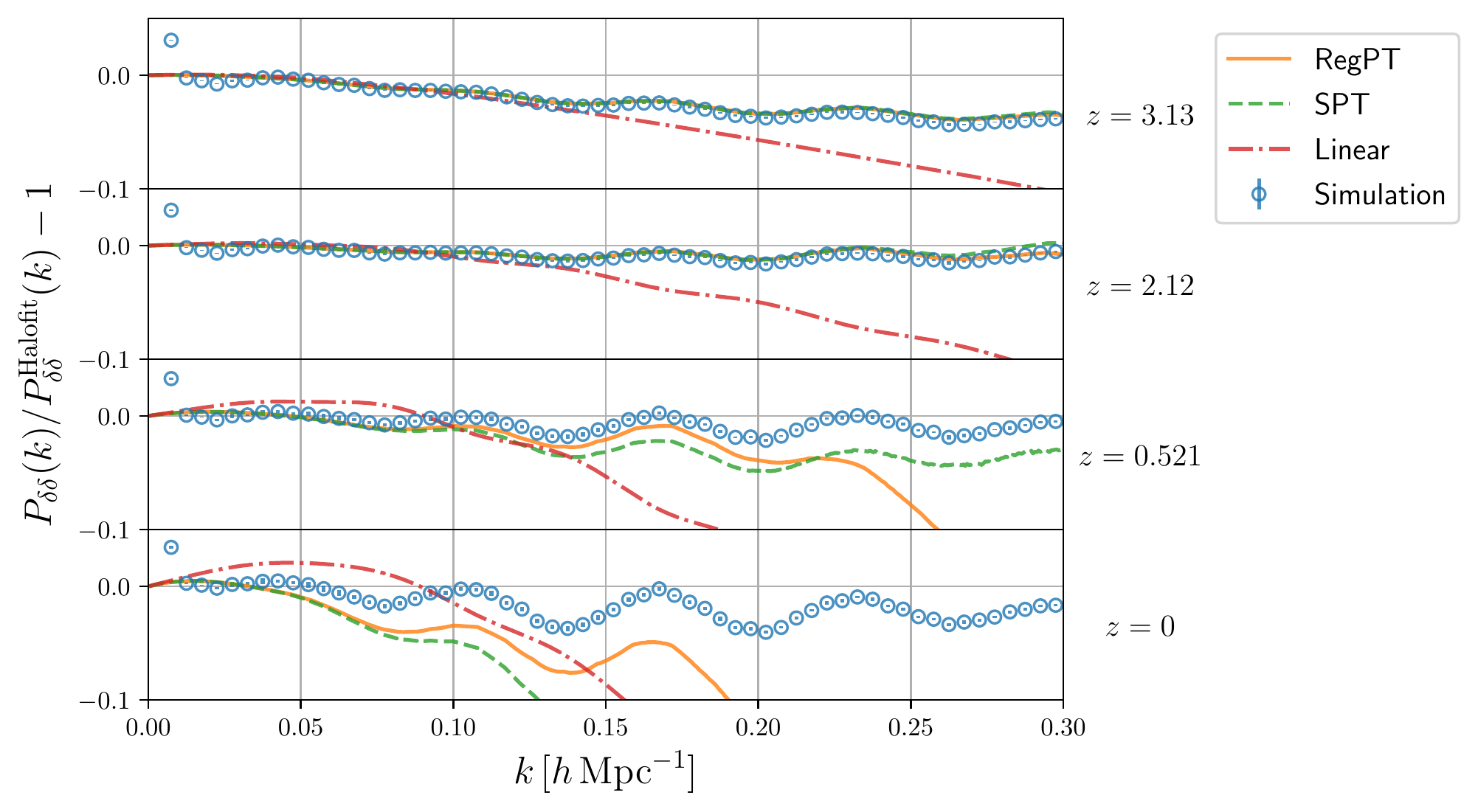}
  \caption{Same as Figure~\ref{fig:Pk} but for $z = 3.13, 2.12, 0.521, 0$.}
  \label{fig:Pk_z}
\end{figure*}

\begin{figure*}
  \includegraphics[width=0.75\columnwidth]{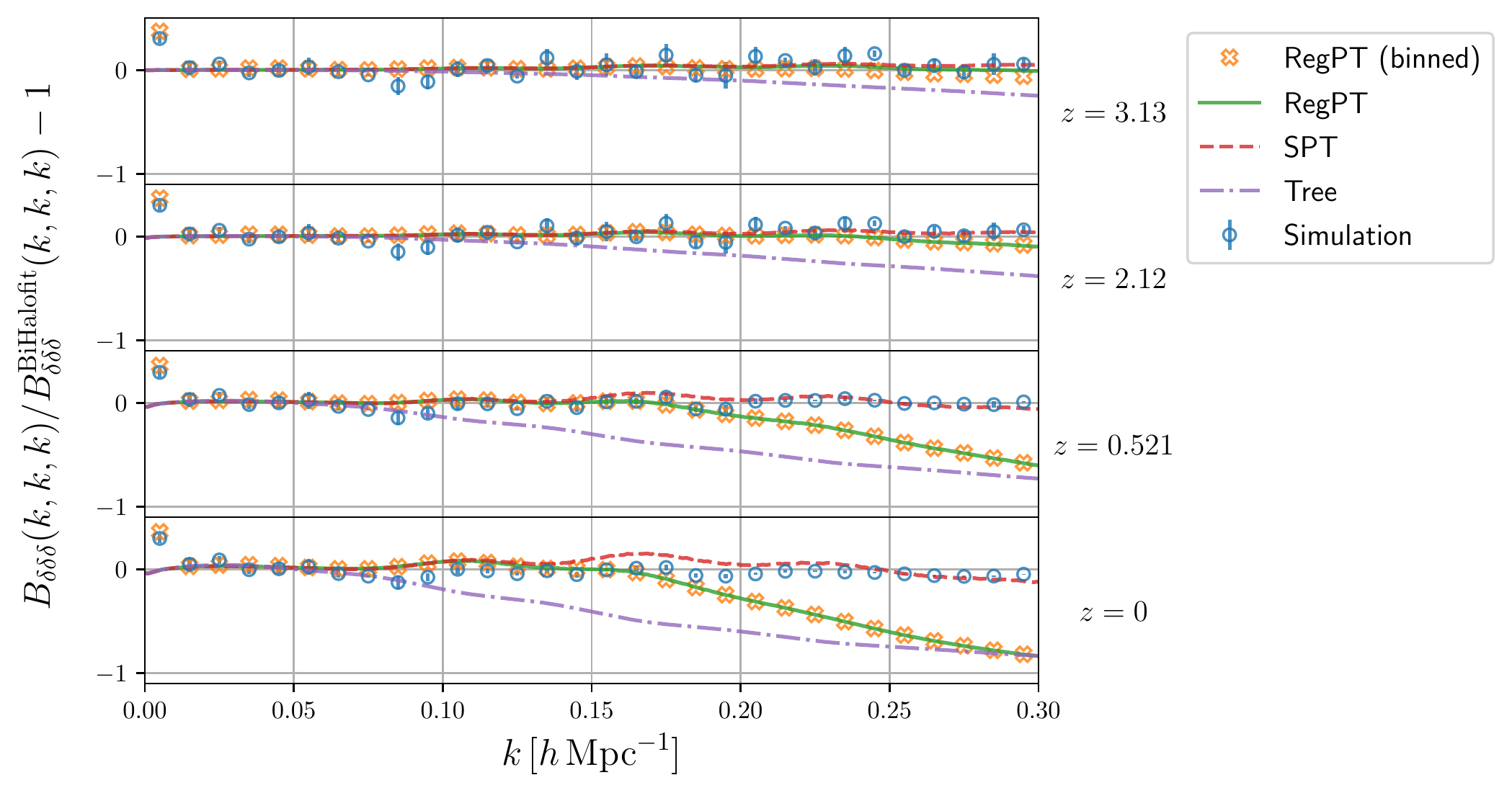}
  \caption{Same as Figure~\ref{fig:Bk_equil} but for $z = 3.13, 2.12, 0.521, 0$.}
  \label{fig:Bk_equil_z}
\end{figure*}

\begin{figure*}
  \includegraphics[width=\textwidth]{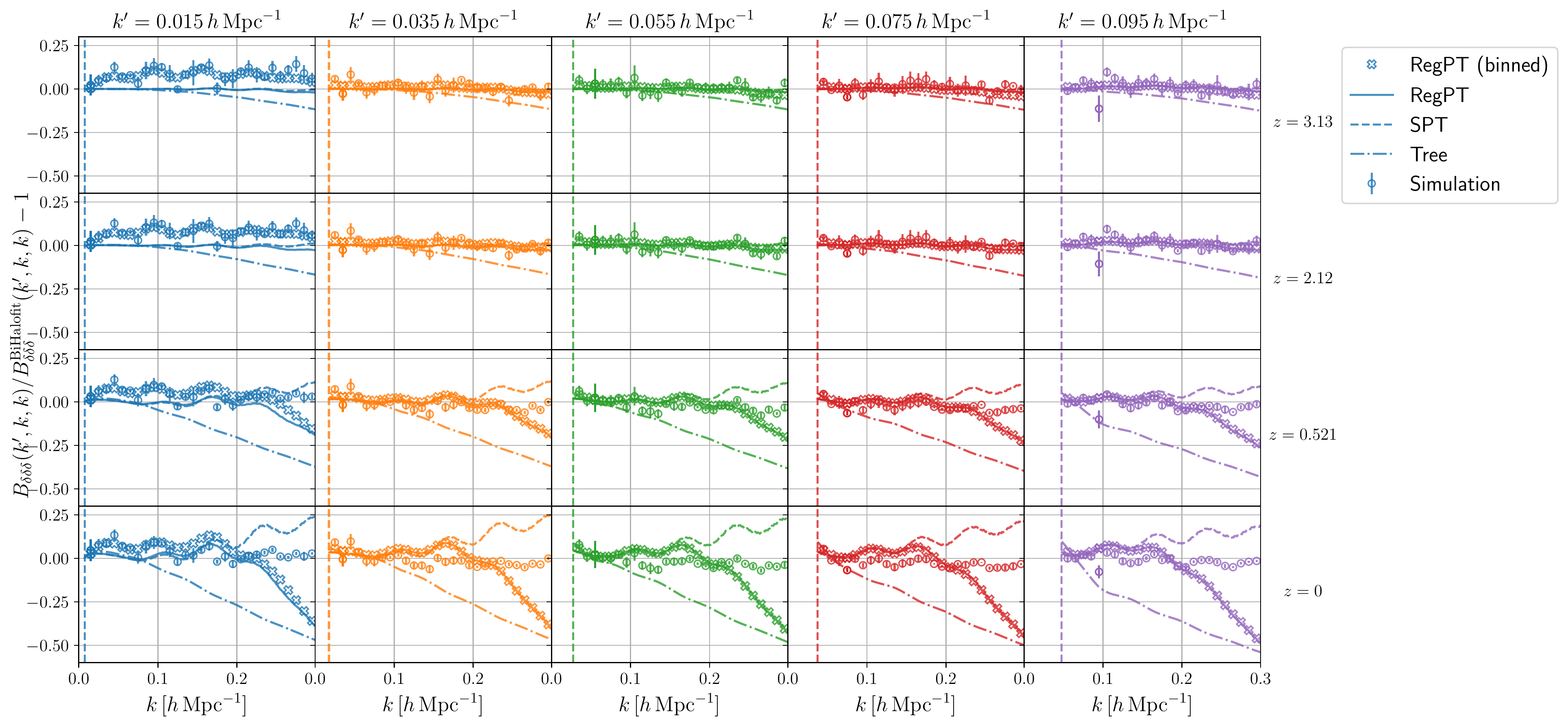}
  \caption{Same as Figure~\ref{fig:Bk_iso1} but for $z = 3.13, 2.12, 0.521, 0$.}
  \label{fig:Bk_iso1_z}
\end{figure*}

\begin{figure*}
  \includegraphics[width=\textwidth]{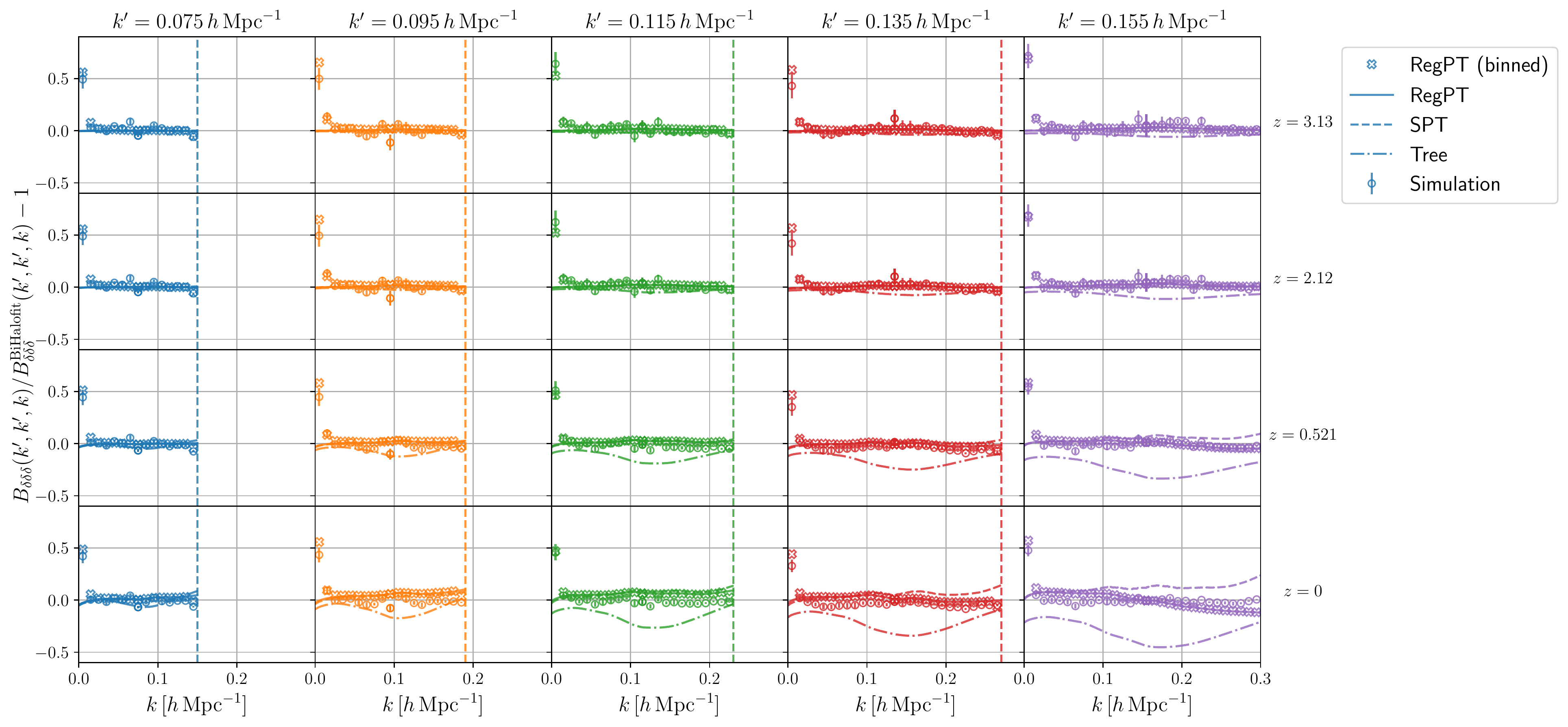}
  \caption{Same as Figure~\ref{fig:Bk_iso2} but for $z = 3.13, 2.12, 0.521, 0$.}
  \label{fig:Bk_iso2_z}
\end{figure*}

In Figures~\ref{fig:Pk_validation_z}, \ref{fig:Bk_equil_validation_z}, \ref{fig:Bk_iso1_validation_z}, and
\ref{fig:Bk_iso2_validation_z}, results of validation for reconstruction at redshifts $z = 3, 2, 0.5, 0$
are shown. Similarly, in Figures~\ref{fig:Pk_z}, \ref{fig:Bk_equil_z}, \ref{fig:Bk_iso1_z}, and
\ref{fig:Bk_iso2_z}, comparison with $N$-body simulations for density power spectra and bispectra
at redshifts $z = 3.13, 2.12, 0.521, 0$ is shown.
The general trend is similar to the results at $z = 1$ for validation
and $z = 0.901$ for comparison with $N$-body simulations but in both cases,
accuracy becomes better at higher redshifts, where non-linearity is not so strong.

% Create the reference section using BibTeX:
\bibliographystyle{apsrev4-2}
\bibliography{main}

\end{document}